\documentclass[aps,prl,twocolumn,superscriptaddress,floatfix,nofootinbib,longbibliography]{revtex4-2}
\usepackage{amsmath}
\usepackage{times}
\usepackage{amssymb}
\usepackage{graphicx}
\usepackage{soul}
\usepackage[colorlinks=true,linkcolor=blue, citecolor=blue, urlcolor=blue, bookmarks]{hyperref}
\makeatletter
\usepackage[T1]{fontenc}
\usepackage{amsmath,amsfonts,amssymb,amsthm,bbm,bm, braket}
\usepackage{wasysym}
\usepackage{comment}
\usepackage{scalerel}
\usepackage{enumerate}
\usepackage{amssymb}
\usepackage{graphicx}
\usepackage{mathtools}
\usepackage{ifthen}
\usepackage{tensor}
\usepackage{tikz}
\usepackage{tikz-network}
\usetikzlibrary{patterns,decorations.pathreplacing}

\usepackage{color}
\usepackage{longtable}
\usepackage[normalem]{ulem} 

\theoremstyle{break}       
\usepackage{color}
\definecolor{myred}{RGB}{232,102,102}
\definecolor{myblue}{RGB}{187,187,255}
\definecolor{myorange0}{RGB}{252,226,5}
\definecolor{myorange0c}{RGB}{255,255,255}
\definecolor{myorange}{RGB}{255,165,0}
\definecolor{mygrey}{RGB}{105,105,105}
\definecolor{OliveGreen}{RGB}{85,107,47}
\definecolor{NavyBlue}{RGB}{50,50,168}
\definecolor{mygreen}{RGB}{34,139,34}
\definecolor{myY}{RGB}{220,255,203}
\definecolor{myYO}{RGB}{255, 220, 151}

\definecolor{mygreenc}{RGB}{150,50,50}
\usepackage{tensor}
\definecolor{mygreenc}{RGB}{150,50,50}
\newcommand{\fourPEPSO}[2]{
\draw[very thick] (#1-0.5,#2)--(#1+0.5,#2);
\draw[very thick] (#1-0.25,#2-0.5)--(#1+0.25,#2+0.5);
\draw[very thick] (#1,#2)--(#1,#2+0.5);
\draw[thick, fill=myorange] (#1,#2) circle (0.25);
}
\newcommand{\fourPEPSDB}[2]{
\draw[very thick] (#1-0.5,#2)--(#1+0.5,#2);
\draw[very thick] (#1-0.25,#2-0.5)--(#1+0.25,#2+0.5);
\draw[very thick] (#1,#2)--(#1,#2+0.5);
\draw[thick, fill=NavyBlue] (#1,#2) circle (0.25);
}

\newcommand{\fourPEPSGR}[2]{
\draw[very thick] (#1-0.5,#2)--(#1+0.5,#2);
\draw[very thick] (#1-0.25,#2-0.5)--(#1+0.25,#2+0.5);
\draw[very thick] (#1,#2)--(#1,#2+0.5);
\draw[thick, fill=mygrey] (#1,#2) circle (0.25);
}

\newcommand{\upPEPSGR}[2]{
\draw[very thick] (#1-0.5,#2)--(#1+0.5,#2);
\draw[very thick] (#1-0.25,#2-0.5)--(#1,#2);
\draw[very thick] (#1,#2) .. controls (#1+0.125,#2+0.25) and (#1+0.25,#2+0.5) .. (#1+0.1,#2+0.65);
\draw[very thick] (#1,#2)--(#1,#2+0.5);
\draw[thick, fill=mygrey] (#1,#2) circle (0.25);
}
\newcommand{\upPEPSGLG}[2]{
\draw[very thick] (#1-0.5,#2)--(#1+0.5,#2);
\draw[very thick] (#1-0.25,#2-0.5)--(#1,#2);
\draw[very thick] (#1,#2) .. controls (#1+0.125,#2+0.25) and (#1+0.25,#2+0.5) .. (#1+0.1,#2+0.65);
\draw[very thick] (#1,#2)--(#1,#2+0.5);
\draw[thick, fill=myY] (#1,#2) circle (0.25);
}
\newcommand{\upPEPSGO}[2]{
\draw[very thick] (#1-0.5,#2)--(#1+0.5,#2);
\draw[very thick] (#1-0.25,#2-0.5)--(#1,#2);
\draw[very thick] (#1,#2) .. controls (#1+0.125,#2+0.25) and (#1+0.25,#2+0.5) .. (#1+0.1,#2+0.65);
\draw[very thick] (#1,#2)--(#1,#2+0.5);
\draw[thick, fill=myorange] (#1,#2) circle (0.25);
}
\newcommand{\downPEPSGR}[2]{
\draw[very thick] (#1-0.5,#2)--(#1+0.5,#2);
\draw[very thick] (#1+0.25,#2+0.5)--(#1,#2);
\draw[very thick] (#1,#2) .. controls (#1-0.125,#2-0.25) and (#1-0.25,#2-0.5) .. (#1-0.35,#2-0.35);
\draw[very thick] (#1,#2)--(#1,#2+0.5);
\draw[thick, fill=mygrey] (#1,#2) circle (0.25);
}
\newcommand{\downPEPSLG}[2]{
\draw[very thick] (#1-0.5,#2)--(#1+0.5,#2);
\draw[very thick] (#1+0.25,#2+0.5)--(#1,#2);
\draw[very thick] (#1,#2) .. controls (#1-0.125,#2-0.25) and (#1-0.25,#2-0.5) .. (#1-0.35,#2-0.35);
\draw[very thick] (#1,#2)--(#1,#2+0.5);
\draw[thick, fill=myY] (#1,#2) circle (0.25);
}
\newcommand{\downPEPSDB}[2]{
\draw[very thick] (#1-0.5,#2)--(#1+0.5,#2);
\draw[very thick] (#1+0.25,#2+0.5)--(#1,#2);
\draw[very thick] (#1,#2) .. controls (#1-0.125,#2-0.25) and (#1-0.25,#2-0.5) .. (#1-0.35,#2-0.35);
\draw[very thick] (#1,#2)--(#1,#2+0.5);
\draw[thick, fill=NavyBlue] (#1,#2) circle (0.25);
}
\newcommand{\leftdownPEPSGR}[2]{
\draw[very thick] (#1,#2)--(#1+0.5,#2);
\draw[very thick] (#1,#2) .. controls (#1-0.45,#2) .. (#1-0.5,#2+0.15);
\draw[very thick] (#1+0.25,#2+0.5)--(#1,#2);
\draw[very thick] (#1,#2) .. controls (#1-0.125,#2-0.25) and (#1-0.25,#2-0.5) .. (#1-0.35,#2-0.35);
\draw[very thick] (#1,#2)--(#1,#2+0.5);
\draw[thick, fill=mygrey] (#1,#2) circle (0.25);
}
\newcommand{\rightdownPEPSGR}[2]{
\draw[very thick] (#1,#2)--(#1-0.5,#2);
\draw[very thick] (#1,#2) .. controls (#1+0.45,#2) .. (#1+0.5,#2+0.15);
\draw[very thick] (#1+0.25,#2+0.5)--(#1,#2);
\draw[very thick] (#1,#2) .. controls (#1-0.125,#2-0.25) and (#1-0.25,#2-0.5) .. (#1-0.35,#2-0.35);
\draw[very thick] (#1,#2)--(#1,#2+0.5);
\draw[thick, fill=mygrey] (#1,#2) circle (0.25);
}
\newcommand{\leftPEPSGR}[2]{
\draw[very thick] (#1,#2)--(#1+0.5,#2);
\draw[very thick] (#1,#2) .. controls (#1-0.45,#2) .. (#1-0.5,#2+0.15);
\draw[very thick] (#1+0.25,#2+0.5)--(#1-0.25,#2-0.5);
\draw[very thick] (#1,#2)--(#1,#2+0.5);
\draw[thick, fill=mygrey] (#1,#2) circle (0.25);
}
\newcommand{\leftPEPSLG}[2]{
\draw[very thick] (#1,#2)--(#1+0.5,#2);
\draw[very thick] (#1,#2) .. controls (#1-0.45,#2) .. (#1-0.5,#2+0.15);
\draw[very thick] (#1+0.25,#2+0.5)--(#1-0.25,#2-0.5);
\draw[very thick] (#1,#2)--(#1,#2+0.5);
\draw[thick, fill=myY] (#1,#2) circle (0.25);
}
\newcommand{\crossdot}[2]
{
\draw[very thick] (#1-0.1,#2-0.1)--(#1+0.1,#2+0.1);
\draw[very thick] (#1-0.1,#2+0.1)--(#1+0.1,#2-0.1);
}
\newcommand{\leftPEPSDB}[2]{
\draw[very thick] (#1,#2)--(#1+0.5,#2);
\draw[very thick] (#1,#2) .. controls (#1-0.45,#2) .. (#1-0.5,#2+0.15);
\draw[very thick] (#1+0.25,#2+0.5)--(#1-0.25,#2-0.5);
\draw[very thick] (#1,#2)--(#1,#2+0.5);
\draw[thick, fill=NavyBlue] (#1,#2) circle (0.25);
}
\newcommand{\rightPEPSGR}[2]{
\draw[very thick] (#1,#2)--(#1-0.5,#2);
\draw[very thick] (#1,#2) .. controls (#1+0.45,#2) .. (#1+0.5,#2+0.15);
\draw[very thick] (#1+0.25,#2+0.5)--(#1-0.25,#2-0.5);
\draw[very thick] (#1,#2)--(#1,#2+0.5);
\draw[thick, fill=mygrey] (#1,#2) circle (0.25);
}
\newcommand{\rightPEPSLG}[2]{
\draw[very thick] (#1,#2)--(#1-0.5,#2);
\draw[very thick] (#1,#2) .. controls (#1+0.45,#2) .. (#1+0.5,#2+0.15);
\draw[very thick] (#1+0.25,#2+0.5)--(#1-0.25,#2-0.5);
\draw[very thick] (#1,#2)--(#1,#2+0.5);
\draw[thick, fill=myY] (#1,#2) circle (0.25);
}
\newcommand{\rightPEPSO}[2]{
\draw[very thick] (#1,#2)--(#1-0.5,#2);
\draw[very thick] (#1,#2) .. controls (#1+0.45,#2) .. (#1+0.5,#2+0.15);
\draw[very thick] (#1+0.25,#2+0.5)--(#1-0.25,#2-0.5);
\draw[very thick] (#1,#2)--(#1,#2+0.5);
\draw[thick, fill=myorange] (#1,#2) circle (0.25);
}
\newcommand{\rightupPEPSGR}[2]{
\draw[very thick] (#1,#2)--(#1-0.5,#2);
\draw[very thick] (#1,#2) .. controls (#1+0.45,#2) .. (#1+0.5,#2+0.15);
\draw[very thick] (#1,#2)--(#1-0.25,#2-0.5);
\draw[very thick] (#1,#2) .. controls (#1+0.125,#2+0.25) and (#1+0.25,#2+0.5) .. (#1+0.1,#2+0.65);
\draw[very thick] (#1,#2)--(#1,#2+0.5);
\draw[thick, fill=mygrey] (#1,#2) circle (0.25);
}
\newcommand{\leftupPEPSGR}[2]{
\draw[very thick] (#1,#2)--(#1+0.5,#2);
\draw[very thick] (#1,#2) .. controls (#1-0.45,#2) .. (#1-0.5,#2+0.15);
\draw[very thick] (#1,#2)--(#1-0.25,#2-0.5);
\draw[very thick] (#1,#2)--(#1,#2+0.5);
\draw[very thick] (#1,#2) .. controls (#1+0.125,#2+0.25) and (#1+0.25,#2+0.5) .. (#1+0.1,#2+0.65);
\draw[thick, fill=mygrey] (#1,#2) circle (0.25);
}
\newcommand{\MYtriangle}[2]{
 \coordinate (Origin) at (#1,#2);
\filldraw [thick, fill=white, even odd rule] ($(Origin)+(-.0cm,{0.666*cos(60)*0.3cm})$) coordinate (Triangle) -- ++(0.15cm,-{cos(60)*0.3cm}) -- ++(-0.3cm,0.0cm) -- ++(0.15cm,{cos(60)*0.3cm}) -- cycle;
}
\newcommand{\MYdiamond}[2]{
\draw[thick, fill=white](#1,#2+0.15)--(#1-0.15,#2)--(#1,#2-0.15)--(#1+0.15,#2)--(#1,#2+0.15);
}
\newcommand{\Wgategreen}[2]{
\draw[very thick] (#1-0.5, #2 +0.5) -- (#1+0.5,#2-0.5);
\draw[very thick] (#1-0.5,#2-0.5) -- (#1+0.5,#2+0.5);
\draw[ thick, fill=mygreen, rounded corners=2pt] (#1-0.25,#2+0.25) rectangle (#1+0.25,#2-0.25);
\draw[thick] (#1,#2+0.15) -- (#1+0.15,#2+0.15) -- (#1+0.15,#2);
}
\newcommand{\Wgateblue}[2]{
\draw[very thick] (#1-0.5, #2 +0.5) -- (#1+0.5,#2-0.5);
\draw[very thick] (#1-0.5,#2-0.5) -- (#1+0.5,#2+0.5);
\draw[ thick, fill=myblue, rounded corners=2pt] (#1-0.25,#2+0.25) rectangle (#1+0.25,#2-0.25);
\draw[thick] (#1,#2+0.15) -- (#1+0.15,#2+0.15) -- (#1+0.15,#2);
}
\newcommand{\MYcircleB}[2]{
\draw[ thick, fill=black] (#1,#2) circle (0.2);
}

\newcommand{\fourPEPSG}[2]{
\draw[very thick] (#1-0.5,#2)--(#1+0.5,#2);
\draw[very thick] (#1-0.25,#2-0.5)--(#1+0.25,#2+0.5);
\draw[very thick] (#1,#2)--(#1,#2+0.5);
\draw[thick, fill=mygreen] (#1,#2) circle (0.25);
}
\newcommand{\threePEPSG}[2]{
\draw[very thick] (#1,#2)--(#1+0.5,#2);
\draw[very thick] (#1-0.25,#2-0.5)--(#1+0.25,#2+0.5);
\draw[very thick] (#1,#2)--(#1,#2+0.5);
\draw[thick, fill=mygreen] (#1,#2) circle (0.25);
}
\newcommand{\threePEPSB}[2]{
\draw[very thick] (#1,#2)--(#1+0.5,#2);
\draw[very thick] (#1-0.25,#2-0.5)--(#1+0.25,#2+0.5);
\draw[very thick] (#1,#2)--(#1,#2+0.5);
\draw[thick, fill=myblue] (#1,#2) circle (0.25);
}
\newcommand{\fourPEPSB}[2]{
\draw[very thick] (#1-0.5,#2)--(#1+0.5,#2);
\draw[very thick] (#1-0.25,#2-0.5)--(#1+0.25,#2+0.5);
\draw[very thick] (#1,#2)--(#1,#2+0.5);
\draw[thick, fill=myblue] (#1,#2) circle (0.25);
}
\newcommand{\midarrow}{\tikz \draw[-triangle 90] (0,0) -- +(.1,0);}
\newcommand{\fourPEPSBDire}[2]{
\begin{scope}[very thick, every node/.style={sloped,allow upside down}]
 \draw (#1-1,#2)--node {\midarrow} (#1,#2);
 \draw (#1+0.25,#2)--node {\midarrow} (#1+1,#2);
 \draw (#1+0.1, #2+0.2)--node {\midarrow}(#1+0.5, #2+1);
 \draw (#1-0.5,#2-1)--node {\midarrow} (#1,#2);
 \draw[very thick] (#1,#2)--(#1,#2+1);
\end{scope}
\draw[thick, fill=myblue] (#1,#2) circle (0.25);
}
\newcommand{\fourPEPSR}[2]{
\draw[very thick] (#1-0.5,#2)--(#1+0.5,#2);
\draw[very thick] (#1-0.25,#2-0.5)--(#1+0.25,#2+0.5);
\draw[very thick] (#1,#2)--(#1,#2+0.5);
\draw[thick, fill=myred] (#1,#2) circle (0.25);
}

\newcommand{\upPEPSG}[2]{
\draw[very thick] (#1-0.5,#2)--(#1+0.5,#2);
\draw[very thick] (#1-0.25,#2-0.5)--(#1,#2);
\draw[very thick] (#1,#2) .. controls (#1+0.125,#2+0.25) and (#1+0.25,#2+0.5) .. (#1+0.1,#2+0.65);
\draw[very thick] (#1,#2)--(#1,#2+0.5);
\draw[thick, fill=mygreen] (#1,#2) circle (0.25);
}
\newcommand{\downPEPSG}[2]{
\draw[very thick] (#1-0.5,#2)--(#1+0.5,#2);
\draw[very thick] (#1+0.25,#2+0.5)--(#1,#2);
\draw[very thick] (#1,#2) .. controls (#1-0.125,#2-0.25) and (#1-0.25,#2-0.5) .. (#1-0.35,#2-0.35);
\draw[very thick] (#1,#2)--(#1,#2+0.5);
\draw[thick, fill=mygreen] (#1,#2) circle (0.25);
}
\newcommand{\leftdownPEPSG}[2]{
\draw[very thick] (#1,#2)--(#1+0.5,#2);
\draw[very thick] (#1,#2) .. controls (#1-0.45,#2) .. (#1-0.5,#2+0.15);
\draw[very thick] (#1+0.25,#2+0.5)--(#1,#2);
\draw[very thick] (#1,#2) .. controls (#1-0.125,#2-0.25) and (#1-0.25,#2-0.5) .. (#1-0.35,#2-0.35);
\draw[very thick] (#1,#2)--(#1,#2+0.5);
\draw[thick, fill=mygreen] (#1,#2) circle (0.25);
}
\newcommand{\rightdownPEPSG}[2]{
\draw[very thick] (#1,#2)--(#1-0.5,#2);
\draw[very thick] (#1,#2) .. controls (#1+0.45,#2) .. (#1+0.5,#2+0.15);
\draw[very thick] (#1+0.25,#2+0.5)--(#1,#2);
\draw[very thick] (#1,#2) .. controls (#1-0.125,#2-0.25) and (#1-0.25,#2-0.5) .. (#1-0.35,#2-0.35);
\draw[very thick] (#1,#2)--(#1,#2+0.5);
\draw[thick, fill=mygreen] (#1,#2) circle (0.25);
}
\newcommand{\leftPEPSG}[2]{
\draw[very thick] (#1,#2)--(#1+0.5,#2);
\draw[very thick] (#1,#2) .. controls (#1-0.45,#2) .. (#1-0.5,#2+0.15);
\draw[very thick] (#1+0.25,#2+0.5)--(#1-0.25,#2-0.5);
\draw[very thick] (#1,#2)--(#1,#2+0.5);
\draw[thick, fill=mygreen] (#1,#2) circle (0.25);
}
\newcommand{\rightPEPSG}[2]{
\draw[very thick] (#1,#2)--(#1-0.5,#2);
\draw[very thick] (#1,#2) .. controls (#1+0.45,#2) .. (#1+0.5,#2+0.15);
\draw[very thick] (#1+0.25,#2+0.5)--(#1-0.25,#2-0.5);
\draw[very thick] (#1,#2)--(#1,#2+0.5);
\draw[thick, fill=mygreen] (#1,#2) circle (0.25);
}
\newcommand{\rightupPEPSG}[2]{
\draw[very thick] (#1,#2)--(#1-0.5,#2);
\draw[very thick] (#1,#2) .. controls (#1+0.45,#2) .. (#1+0.5,#2+0.15);
\draw[very thick] (#1,#2)--(#1-0.25,#2-0.5);
\draw[very thick] (#1,#2) .. controls (#1+0.125,#2+0.25) and (#1+0.25,#2+0.5) .. (#1+0.1,#2+0.65);
\draw[very thick] (#1,#2)--(#1,#2+0.5);
\draw[thick, fill=mygreen] (#1,#2) circle (0.25);
}
\newcommand{\leftupPEPSG}[2]{
\draw[very thick] (#1,#2)--(#1+0.5,#2);
\draw[very thick] (#1,#2) .. controls (#1-0.45,#2) .. (#1-0.5,#2+0.15);
\draw[very thick] (#1,#2)--(#1-0.25,#2-0.5);
\draw[very thick] (#1,#2)--(#1,#2+0.5);
\draw[very thick] (#1,#2) .. controls (#1+0.125,#2+0.25) and (#1+0.25,#2+0.5) .. (#1+0.1,#2+0.65);
\draw[thick, fill=mygreen] (#1,#2) circle (0.25);
}

\newcommand{\MYsquareW}[2]{
\draw[ thick, fill=white,rounded corners=0pt] (#1-0.1,#2+0.1) rectangle (#1+0.1,#2-0.1);
}

\newcommand{\MYcircle}[2]{
\draw[thick, fill=white] (#1,#2) circle (0.1cm); }
\usepackage{graphicx} 

\makeatother

\usepackage{xr-hyper}
\usepackage[colorlinks=true,linkcolor=blue, citecolor=blue, urlcolor=blue, bookmarks]{hyperref}

\hypersetup{
	pdftitle={Dual-isometric Projected Entangled Pair States},
	pdfsubject={Many-body quantum physics},
	pdfauthor={Xie-Hang Yu, J. Ignacio Cirac, Pavel Kos, Georgios Styliaris},
	pdfkeywords={PEPS, MPS, quantum states}
}

\begin{document}
\title{Dual-isometric Projected Entangled Pair States}
\author{Xie-Hang Yu}
\author{J. Ignacio Cirac}
\author{Pavel Kos}
\thanks{These authors contributed equally.}
\author{Georgios Styliaris}
\thanks{These authors contributed equally.}
\affiliation{%
Max-Planck-Institut f{\"{u}}r Quantenoptik, Hans-Kopfermann-Str. 1, 85748 Garching, Germany
}%
\affiliation{%
Munich Center for Quantum Science and Technology (MCQST), Schellingstr. 4, 80799 M{\"{u}}nchen, Germany
}%
\begin{abstract}
Efficient characterization of higher dimensional many-body physical states presents significant challenges. In this paper, we propose a new class of Projected Entangled Pair States (PEPS) that incorporates two isometric conditions. This new class facilitates the efficient calculation of general local observables and certain two-point correlation functions, which have been previously shown to be intractable for general PEPS, or PEPS with only a single isometric constraint. Despite incorporating two isometric conditions, our class preserves the rich physical structure while enhancing the analytical capabilities. It features a large set of tunable parameters, with only a subleading correction compared to that of general PEPS. Furthermore, we analytically demonstrate that this class can encode universal quantum computation and can represent a transition from topological to trivial order.
\end{abstract}
\maketitle

\emph{Introduction.---}Efficiently representing strongly correlated many-body quantum states and extracting their relevant physical properties remain a pivotal challenge in dimensions higher than one. Tensor network methods offer an approach to address this problem. Following the successful application of 1D matrix product states (MPS)~\cite{fannes1992finitely,cirac2021matrix,Haegeman_Verstraete_2017},
Ref.~\cite{verstraete2004renormalization} introduced their higher-dimensional generalization, projected entangled pair states (PEPS). These states satisfy an entanglement area law~\cite{RevModPhys.82.277} and are considered a robust ansatz for representing ground states of gapped local Hamiltonians~\cite{Niggemann_Klümper_Zittartz_1997,verstraete2006matrix,verstraete2006criticality,hastings2007area,brandao2015exponential,molnar2015approximating,Anshu_Harrow_Soleimanifar_2022,Anshu_Arad_Gosset_2022,ORourke_Chan_2023}.
PEPS also play an important role in quantum dynamics~\cite{PhysRevA.83.052321,PhysRevB.106.195132,10.21468/SciPostPhys.15.4.158}, statistical mechanics~\cite{PhysRevLett.99.120601,PhysRevLett.103.160601,PhysRevB.81.174411,PhysRevB.93.125115}, the classification of phases~\cite{chen2011classification,PhysRevB.84.165139,PhysRevLett.111.090501,PhysRevX.8.031031}, and quantum machine learning~\cite{liu2023theory,PhysRevB.103.125117,Azizi2020LearningPT}.

Despite their advantages, finite PEPS still suffer from the rapid growth of the computational resources required for the calculation of physical quantities, such as expectation values of local observables.
In practice, even approximate algorithms on finite PEPS are costly and the errors are often hard to control~\cite{PhysRevB.90.064425,Lubasch_2014,PhysRevLett.125.210504}. This limitation also holds for translationally-invariant systems~\cite{vanderstraeten2016gradient}. It has been shown that the exact contraction of a general PEPS is \#P-hard~\cite{PhysRevLett.98.140506}, even for typical instances~\cite{PhysRevResearch.2.013010}, which serves as a fundamental limitation.

One way to address this challenge is to consider some subclass of PEPS. One such class is the isometric PEPS (iso-PEPS)~\cite{haghshenas2019conversion,PhysRevLett.124.037201}, which extends the canonical form of 1D MPS to higher dimensions. Their isometric nature allows for the preparation of all iso-PEPS through sequential unitary circuits~\cite{PhysRevLett.128.010607}. This sequential generation defines a time axis in the iso-PEPS and thus facilitates the backward contraction along this time direction~\cite{PhysRevLett.124.037201}. 
Despite being typically short-range correlated~\cite{haag2023typical}, the iso-PEPS family can capture states with complex correlations, such as  topological models and the associated phase transitions~\cite{PhysRevB.101.085117,liu2024topological}.

However, the computation of the local observables for iso-PEPS is not always efficient.
Indeed, this task in 2D is shown to be $\textsf{BQP}$-complete~\cite{malz2024computational}, which, subject to standard complexity theory assumptions, indicates that it is hard to simulate classically. 
Although in principle, some classical algorithms, like the Moses move~\cite{PhysRevLett.124.037201,PhysRevB.107.245118}, can be used to approximate the result, the errors may remain uncontrolled~\cite{PhysRevB.106.245102}.
This restricts the practical utility of iso-PEPS and motivates the search for new classes of contractible PEPS.

\begin{figure}[t]
\includegraphics[width=0.8\columnwidth]{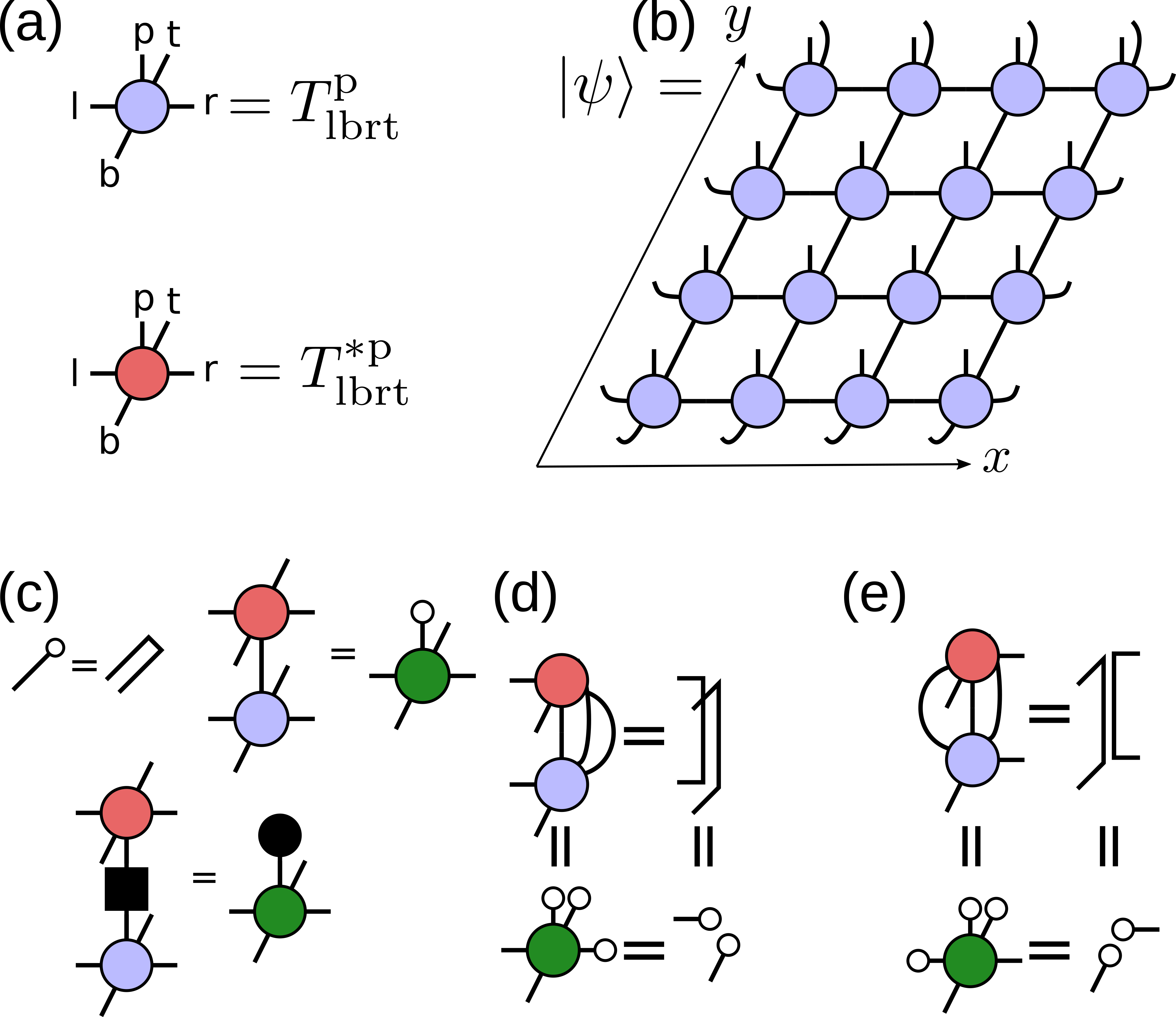}
\caption{a)
Rank-5 tensor $T$ of a PEPS.
The complex conjugate of $T$ is denoted by $T^*$.  b) The
physical state from the contracted local PEPS tensors, with the virtual space mapped to the physical one on the boundary. 
c) 
The local contraction of bra and ket PEPS states, with its folded notation shown on the right.
The bottom panel shows the local expectation value of an observable (black square).  The black dot on the right-hand side represents its vectorized form. 
d) Isometric condition in the unfolded and folded picture. e) Dual isometric condition.}
\label{fig1}
\end{figure}

In this letter, we introduce dual-isometric PEPS (DI-PEPS), a new subclass of iso-PEPS that enhances the computational tractability of tensor networks in higher dimensions, especially  for computing local observables and certain two-point correlation functions. The class is defined by imposing an additional isometric condition onto iso-PEPS, which reduces the calculation of the above quantities in the 2D tensor network to a 1D tensor network. The latter is manageable and can be efficiently computed. More broadly, the dual-isometric condition is analogous to dual-unitary gates~\cite{bertini2019exact}, a class of exactly solvable quantum circuits, and can be seen as a natural extension of these ideas to PEPS.
Furthermore, our work reveals that the DI-PEPS preserve the rich physical structure of iso-PEPS despite the additional constraint. For this, we show that DI-PEPS can encode universal quantum computations after post-selection. As a result, their output probability cannot be efficiently sampled on a classical computer. 
Additionally, by counting the number of free parameters of
DI-PEPS, we show that this number is only reduced at subleading order as compared to conventional PEPS.
Lastly, we develop a class of DI-PEPS which exhibits topological order and has a nontrivial transition to decoupled 1D chains. These findings underscore the DI-PEPS as a potent framework for investigating quantum many-body physics, offering new avenues for understanding and manipulating complex quantum systems.

\emph{PEPS and the folded picture.---}We consider PEPS specified by a rank-$5$ tensor $T_{\mathrm{lbrt}}^{\mathrm{p}}$ for each vertex $(x,y)$ of a 2D square lattice. The index $\mathrm{p}$ represents the physical degree of freedom
with dimension $d$, and $\mathrm{l,b,r,t}$ (left, bottom, right, top) label the virtual degrees of
freedom, each $\chi$-dimensional [Fig.~(\ref{fig1}a)]. Here  $x\in\{1,\cdots,N\}$ and $y\in\{1,\cdots,M\}$.
The wave function is obtained by locally contracting all the virtual
degrees of freedom [Fig.~(\ref{fig1}b)]. At the boundary, the virtual space is isometrically mapped to the physical space; we explain this choice later. Thus, there are in total $(N+2)\times(M+2)-4$ physical sites.

Our setting and results become more transparent via vectorizing; in graphical notation, this corresponds to ``folding'' [Fig.~(\ref{fig1}c)]. Given a fixed computational basis, an observable is vectorized to $\bra{O}$ via $\sum_{ij} O_{ij} \ket{i}\!\bra{j} \mapsto \sum_{ij} O_{ij} \bra{i}\bra{j}$. States are similarly mapped to a bipartite vector such that
\begin{equation}
\braket{\psi|\hat{O}|\psi}=\braket{O|(|\psi^*}\otimes\ket{\psi}).
\end{equation}
Graphically, $T^{*}$ is folded in front of $T$, thereby forming $T^{*}\otimes T = \begin{tikzpicture}[baseline={([yshift=-0.65ex] current bounding box.center)}, scale=0.6]
\fourPEPSG{0}{0}
\end{tikzpicture}
=
\begin{tikzpicture}[baseline={([yshift=-0.4ex] current bounding box.center)}, scale=0.5]
\fourPEPSB{0.35}{0.35}
\fourPEPSR{0}{0}
\end{tikzpicture}$.

\emph{Iso-PEPS.---}Here we briefly review iso-PEPS~\cite{PhysRevLett.124.037201}, which is a subclass of PEPS satisfying the isometric condition 
\begin{equation}
\sum_{\mathrm{r},\mathrm{t},\mathrm{p}}T_{\mathrm{l}_{1}\mathrm{b}_{1}\mathrm{r}\mathrm{t}}^{\mathrm{p}}T_{\mathrm{l}_{2}\mathrm{b}_{2}\mathrm{r}\mathrm{t}}^{*\mathrm{p}}=\delta_{\mathrm{l}_{1},\mathrm{l}_{2}}\delta_{\mathrm{b}_{1},\mathrm{b}_{2}},\label{eq:ISOTNS}
\end{equation}
shown also graphically in Fig.~(\ref{fig1}d). This class can be understood as the 2D analog of the canonical form in MPS. Eq.~\eqref{eq:ISOTNS} directly implies that iso-PEPS can be contracted starting from the top-right direction until an observable is met~\cite{PhysRevLett.124.037201}.

Physically, every iso-PEPS tensor can be sequentially generated in depth $\mathcal O(N+M)$ via a unitary gate ~\cite{PhysRevLett.128.010607}
\begin{equation}
U_{\mathrm{lb}\ket{0}}^{\mathrm{prt}}=T_{\mathrm{lbrt}}^{\mathrm{p}} \label{eq:equivalent_gate}
\end{equation}
with one of its inputs initialized to $\ket{0}$. 
Calculating expectation values of local observables in iso-PEPS is as hard as quantum computation since an expectation value of a bulk operator may correspond to a late-time expectation value in the associated sequential circuit; indeed, this task was recently shown to be $\textsf{BQP}$-complete~\cite{malz2024computational}. Note that the isometric property implies that only a local observable  near the bottom/left boundary of the PEPS can be efficiently calculated~\cite{PhysRevLett.124.037201}. In the sequential quantum circuit picture, this only involves a contribution from the gates on the light cone~\cite{PhysRevLett.128.010607}.

\emph{Dual-isometric PEPS.---}In this letter, we introduce
a new subclass of iso-PEPS called dual-isometric PEPS (DI-PEPS). This is done by requiring another (dual) isometric condition
\begin{equation}
\sum_{\mathrm{l},\mathrm{t},\mathrm{p}}T_{\mathrm{l}\mathrm{b}_{1}\mathrm{r}_{1}\mathrm{t}}^{\mathrm{p}}T_{\mathrm{l}\mathrm{b}_{2}\mathrm{r}_{2}\mathrm{t}}^{*\mathrm{p}}=\delta_{\mathrm{r}_{1},\mathrm{r}_{2}}\delta_{\mathrm{b}_{1},\mathrm{b}_{2}},\label{eq:Dual_ISO_TNS}
\end{equation}
as shown in Fig.~(\ref{fig1}e).
This is in analogy to dual-unitary circuits, where each gate can be interpreted as a valid evolution along a second ``time'' direction. DI-PEPS thus admit an additional sequential preparation direction.
Although we primarily focus on this class, the dual condition of Eq.~\eqref{eq:Dual_ISO_TNS} can be generalized to
\begin{equation}
\sum_{\mathrm{l}_1,\mathrm{l}_2,\mathrm{t},\mathrm{p}}T_{\mathrm{l}_1\mathrm{b}_{1}\mathrm{r}_{1}\mathrm{t}}^{\mathrm{p}}S_{\mathrm{l}_1,\mathrm{l}_2}T_{\mathrm{l}_2\mathrm{b}_{2}\mathrm{r}_{2}\mathrm{t}}^{*\mathrm{p}}=S_{\mathrm{r}_{1},\mathrm{r}_{2}}\delta_{\mathrm{b}_{1},\mathrm{b}_{2}},\label{eq:GDual_ISO_TNS}
\end{equation}
where S is a positive semidefinite matrix (see~\cite{SM}). We refer to this enlarged class as \emph{generalized DI-PEPS}.

The key point is that the dual condition, together with Eq.~\eqref{eq:ISOTNS}, allows an efficient calculation of expectation values of local observables.
This becomes transparent in the folded picture, where the expectation value is expressed as
\begin{equation}
\braket{\psi|O|\psi}=
\begin{tikzpicture}[baseline=(current  bounding  box.center), scale=0.55]
\foreach \i in {0,1,2}
{
\foreach \j in {0,1,2}
{
\fourPEPSG{\i+0.5*\j}{\j}
\MYcircle{\i+0.5*\j}{\j+0.5}
}
\leftPEPSG{-1+\i*0.5}{\i}
\rightPEPSG{3+\i*0.5}{\i}
\downPEPSG{\i-0.5}{-1}
\upPEPSG{\i+1.5}{3}
\MYcircle{-1+\i*0.5}{\i+0.5}
\MYcircle{3+\i*0.5}{\i+0.5}
\MYcircle{\i-0.5}{-0.5}
\MYcircle{\i+1.5}{3.5}
}
\leftdownPEPSG{-1.5}{-1}
\rightdownPEPSG{2.5}{-1}
\leftupPEPSG{0.5}{3}
\rightupPEPSG{4.5}{3}
\foreach \i in {0,1,2,3,4}
{
\MYcircle{0.65+\i}{3.65}
\MYcircle{-1.85+\i}{-1.35}
\MYcircle{-2+\i*0.5}{-0.85+\i}
\MYcircle{3+0.5*\i}{-0.85+\i}
}
\MYcircle{-1.5}{-0.5}
\MYcircle{2.5}{-0.5}
\MYcircle{0.5}{3.5}
\MYcircle{4.5}{3.5}
\MYcircleB{2}{2.6}
\end{tikzpicture}.
\end{equation}
Starting from the top-left corner, we can use Eq.~\eqref{eq:Dual_ISO_TNS} [Fig.~(\ref{fig1}e)] to simplify
the configuration as ${
\begin{tikzpicture}[baseline=(current  bounding  box.center), scale=0.4]
\leftupPEPSG{0}{0}
\upPEPSG{1}{0}
\leftPEPSG{-0.5}{-1}
\fourPEPSG{0.5}{-1}
\MYcircle{0}{0.5}
\MYcircle{1}{0.5}
\MYcircle{-0.5}{0.15}
\MYcircle{0.15}{0.65}
\MYcircle{1.15}{0.65}
\MYcircle{-0.5}{-0.5}
\MYcircle{0.5}{-0.5}
\MYcircle{-1}{-0.85}
\end{tikzpicture}
=
\begin{tikzpicture}[baseline=(current  bounding  box.center), scale=0.4]
\upPEPSG{1}{0}
\leftPEPSG{-0.5}{-1}
\fourPEPSG{0.5}{-1}
\MYcircle{1}{0.5}
\MYcircle{1.15}{0.65}
\MYcircle{-0.5}{-0.5}
\MYcircle{0.5}{-0.5}
\MYcircle{-0.25}{-0.5}
\MYcircle{0.5}{0}
\MYcircle{-1}{-0.85}
\end{tikzpicture}.}$
This procedure can be iterated, as now the same equation can be used in the next row and column. Following this simplification procedure, one finds that all the tensors on the left and right sides of the observable simplify by Eqs.~\eqref{eq:Dual_ISO_TNS} and~\eqref{eq:ISOTNS}, respectively, resulting in
\begin{equation}
\braket{\psi|O|\psi}=
\begin{tikzpicture}[baseline=(current  bounding  box.center), scale=0.55]
\downPEPSG{-0.5}{-1}
\foreach \i in {0,1,2}
{\fourPEPSG{\i*0.5}{\i}}
\foreach \i in {-1,0,1,2}
{
\MYcircle{\i*0.5}{\i+0.5}
\MYcircle{-0.5+\i*0.5}{\i}
\MYcircle{\i*0.5+0.5}{\i}
}
\MYcircle{1.25}{2.5}
\MYcircle{-0.85}{-1.35}
\MYcircleB{1}{2.6}
\end{tikzpicture}.\label{eq:1Dchannel_for_local}
\end{equation}
This is just a 1D tensor network, which
can be efficiently contracted in $\mathcal{O}(M)$. This result also holds for local operators with support larger than $1$. In that case,
Eq.~\eqref{eq:1Dchannel_for_local} includes more than one column but with constant (i.e., system-size independent) width.

Our next point of interest is two-point correlation functions, which play a key role in characterizing many-body properties.
They can be graphically expressed as
\begin{equation}
\braket{\psi|O_1O_2|\psi}=
\begin{tikzpicture}[baseline=(current  bounding  box.center), scale=0.55]
\foreach \i in {0,1,2}
{
\foreach \j in {0,1,2}
{
\fourPEPSG{\i+0.5*\j}{\j}
\MYcircle{\i+0.5*\j}{\j+0.5}
}
\leftPEPSG{-1+\i*0.5}{\i}
\rightPEPSG{3+\i*0.5}{\i}
\downPEPSG{\i-0.5}{-1}
\upPEPSG{\i+1.5}{3}
\MYcircle{-1+\i*0.5}{\i+0.5}
\MYcircle{3+\i*0.5}{\i+0.5}
\MYcircle{\i-0.5}{-0.5}
\MYcircle{\i+1.5}{3.5}
}

\leftdownPEPSG{-1.5}{-1}
\rightdownPEPSG{2.5}{-1}
\leftupPEPSG{0.5}{3}
\rightupPEPSG{4.5}{3}
\foreach \i in {0,1,2,3,4}
{
\MYcircle{0.65+\i}{3.65}
\MYcircle{-1.85+\i}{-1.35}
\MYcircle{-2+\i*0.5}{-0.85+\i}
\MYcircle{3+0.5*\i}{-0.85+\i}
}
\MYcircle{-1.5}{-0.5}
\MYcircle{2.5}{-0.5}
\MYcircle{0.5}{3.5}
\MYcircle{4.5}{3.5}
\MYcircleB{3}{2.6}
\MYcircleB{0}{0.6}
\end{tikzpicture}
.
\label{eq:2pt}
\end{equation}
With the isometric and dual-isometric conditions, one can simplify the diagram from the top-left
and top-right corners similarly as before. The final result is 
\begin{equation}
\braket{\psi|O_1O_2|\psi}=
\begin{tikzpicture}[baseline=(current  bounding  box.center), scale=0.55]
\foreach \i in {0,1,2}
{
\downPEPSG{\i}{0}
\fourPEPSG{\i+0.5}{1}
\fourPEPSG{2.5+\i*0.5}{\i+1}
}
\foreach \i in {0,1,2}
{
\MYcircle{\i}{0.5}
\MYcircle{\i+0.5}{1.5}
\MYcircle{2.5+\i*0.5}{\i+1.5}
\MYcircle{\i-0.35}{-0.35}
\MYcircle{2.5+\i*0.5}{\i}
}
\MYcircle{-0.5}{0}
\MYcircle{0}{1}
\MYcircle{0.75}{1.5}
\MYcircle{1.75}{1.5}
\MYcircle{2.5}{2}
\MYcircle{3}{3}
\MYcircle{4}{3}
\MYcircle{3.75}{3.5}
\MYcircleB{0.5}{1.6}
\MYcircleB{3.5}{3.6}
\foreach \i in {-0.25,2.25}
{
\draw[thick,dashed](\i,-0.75)--(\i,-1.25);
}
\Text[x=1,y=-1]{\scriptsize{$|x_1\!-\!x_2|$}}
\draw[->](0.15,-1)--(-0.17,-1);
\draw[->](1.85,-1)--(2.17,-1);
\foreach \i in {-0.25,1.25}
{
\draw[thick, dashed] (3.5,\i)--(4.5,\i);
}
\Text[x=4,y=0.5]{\scriptsize{$\min\{y_1,y_2\}$}}
\draw[->](4,0.67)--(4,1.17);
\draw[->](4,0.3)--(4,-0.17);
\end{tikzpicture}.
\end{equation}
Here we assume that the two operators
are located at $(x_{1},y_{1})$ and $(x_{2},y_{2})$, 
respectively. 
The reduced 2D part of the tensor network is of size $t_{1}\times t_{2}$,
with $t_{1}=|x_{2}-x_{1}|$ and $t_{2}=\min\{y_{1},y_{2}\}$. 
If either $t_1$ or $t_2$ is constant, the correlator still reduces to an efficiently contractible
1D tensor network with a possibly increased, but constant, bond dimension.
Similar as above, the generalized DI-PEPS~\eqref{eq:GDual_ISO_TNS} preserve the solvability of 1- and 2-point correlations, see~\cite{SM}.


Not only in our case but also in general, the calculation of local or multi-point expectation values in 2D PEPS can be interpreted
as a circuit of 1+1D completely positive maps acting over the virtual space; in general, however, these are not trace-preserving. 
This is established by interpreting tensor-network diagrams, such as the one in Eq.~\eqref{eq:2pt}, along a $45^\circ$ counterclockwise rotated direction, and defining Kraus operators (indexed by $\mathrm{p}$) as $E_{\mathrm{tr,lb}}^{\mathrm{\mathrm{p}}}=T_{\mathrm{l}\mathrm{b}\mathrm{r}\mathrm{t}}^{\mathrm{p}}$.
Pictorially, this can be expressed as 
\begin{equation}
\begin{tikzpicture}[baseline=(current  bounding  box.center), scale=0.55]
\fourPEPSG{0}{0}
\Text[x=-0.67,y=0]{$\mathrm l$}
\Text[x=0.67,y=0]{$\mathrm r$}
\Text[x=-0.33,y=-0.67]{$\mathrm b$}
\Text[x=0.33,y=0.7]{$\mathrm t$}
\MYcircle{0}{0.5}
\end{tikzpicture}
=
\begin{tikzpicture}[baseline=(current  bounding  box.center), scale=0.55]
\Wgategreen{0}{0}
\Text[x=-0.65,y=-0.65]{$\mathrm l$}
\Text[x=0.65,y=-0.65]{$\mathrm b$}
\Text[x=-0.65,y=0.65]{$\mathrm t$}
\Text[x=0.65,y=0.65]{$\mathrm r$}
\end{tikzpicture}, \label{eq:PEPS_AS_QC}
\end{equation}
where the right side depicts the completely positive map in the folded picture.
For the case of iso-PEPS, this map is in addition trace-preserving, i.e., it corresponds to a physical evolution. 
From this point of view, the solvability of DI-PEPS can be connected with the simplifying properties of two-unital space channels from Ref.~\cite{Kos2023circuitsofspacetime}. 

The boundary condition of Fig.~(\ref{fig1}b) also obtains a physical interpretation at this stage. Viewing the PEPS as a quantum circuit, the boundary conditions correspond to initially preparing $N+M$ EPR-pairs, and inputting half of each pair to the circuit.

\emph{Examples of DI-PEPS.---}
Firstly, a known special subfamily of DI-PEPS is
perfect tensors~\cite{pastawski2015holographic,evenbly2017hyperinvariant,steinberg2022conformal}, 
and their generalizations~\cite{steinberg2023holographic},
originally developed in the context of the AdS/CFT correspondence.
However, this subfamily requires an isometric condition under any bipartition, which is a much stronger constraint than that of generic DI-PEPS. 
Secondly, Sequentially Generated States (SGS)~\cite{PhysRevA.77.052306} are also a subfamily of  generalized DI-PEPS from Eq.~\eqref{eq:GDual_ISO_TNS} (see~\cite{SM}).

As mentioned in Eq.~(\ref{eq:equivalent_gate}), DI-PEPS can be prepared with a sequential quantum circuit with the additional condition
$\sum_{\mathrm{ptl}}U_{\mathrm{lb}_1\ket{0}}^{\mathrm{pr}_1\mathrm{t}}
(U_{\mathrm{lb}_2\ket{0}}^{\mathrm{pr}_2\mathrm{t}})^*=\delta_{\mathrm{b}_1,\mathrm{b}_2}\delta_{\mathrm{r}_1,\mathrm{r}_2}$,
pictorially expressed as
$
\begin{tikzpicture}[baseline=(current bounding box.center), scale=0.50]
\draw[very thick] (-0.5,  +0.5) -- (+0.75,-0.5);
\draw[very thick] (-0.5,-0.5) -- (+0.75,+0.5);
\draw[very thick] (0.125,0.5) -- (0.125,-0.5);
\draw[ thick, fill=mygreen, rounded corners=2pt] (-0.25,0.25) rectangle (0.5,-0.25);
\draw[thick] (0.25,+0.15) -- (+0.4,+0.15) -- (+0.4,0);
\Text[x=1.1,y=-0.55]{\scriptsize $\ket{0}$};
\MYcircle{-0.5}{0.5}
\MYcircle{-0.5}{-0.5}
\MYcircle{0.75}{0.5}
\end{tikzpicture}
=
\begin{tikzpicture}[baseline=(current bounding box.center), scale=0.50]
\draw[very thick](0,0.25)--(0,0.75);
\draw[very thick](0,-0.25)--(0,-0.75);
\MYcircle{0}{0.25}
\MYcircle{0}{-0.25}
\end{tikzpicture}
$.
Some classes of unitary gates satisfying the above are:

\emph{Permutation-phase gates:}
$U=P_{231}D$ composed with arbitrary single site gates, where $P_{231}$ is the shift-permutation $P_{231}=
\begin{tikzpicture}[baseline=(current bounding box.center), scale=0.50]
\draw[very thick](-0.5,-0.5)--(0,0.5);
\draw[very thick](0,-0.5)--(0.5,0.5);
\draw[very thick](0.5,-0.5)--(-0.5,0.5);
\end{tikzpicture}
$ 
and $D$ is an arbitrary diagonal gate in the computational basis. This family generalizes 
dual-unitary gates, for which $P_{21}$ is just the SWAP~\cite{bertini2019exact}, to a tripartite setting.

{\emph{3-qubit gates ($d=\chi=2$):}
We take a non-exhaustive ansatz 
$U=\prod_{\alpha}\exp(iQ_{\alpha}\sigma^{\alpha}_{2}\sigma^{\alpha}_{3})\prod_{\beta}\exp(iJ_{\beta}\sigma^{\beta}_{1}\sigma^{\beta}_{2})$,
where $\alpha, \beta$ run over $1,2,3$ and $\sigma_{m}^{\alpha}$ is the $\alpha$
Pauli matrix acting on the $m$-th qubit.
We find that $Q_{1}=0, Q_{2}=\pi/4, J_{3}=\pi/4$ or $Q_{2}=\pi/4, \cos{2J_{1}}\cos{2Q_{3}}=\cos{2J_{2}}\cos{2Q_{3}}=0$
satisfies the condition.
Additional arbitrary single site unitaries on all five legs but the ancillary one associated with $\ket{0}$ are allowed.

\emph{Controlled-dual unitaries:}
$U_{\mathrm{lba}}^{\mathrm{prt}}=\frac{1}{\sqrt{d}}\sum_{i}\ket{i}_{\mathrm{p}}\bra{i}_{\mathrm{a}}V_{\mathrm{tr},\mathrm{lb}}^{i}$, composed with arbitrary single site gates,
where $V^{i}$ is a dual-unitary gate~\cite{bertini2019exact}, i.e., it is unitary and also satisfies $\sum_{\mathrm{tl}}V_{\mathrm{tr_{1}},\mathrm{lb_{1}}}^{i}V_{\mathrm{tr_{2}},\mathrm{lb_{2}}}^{*i}=\delta_{\mathrm{r}_{1},\mathrm{r}_{2}}\delta_{\mathrm{b}_{1},\mathrm{b}_{2}}$. More generally, any triunitary quantum gate~\cite{jonay2021triunitary} and multi-directional unitary gate (after grouping some indices)~\cite{mestyan2024multi} is an example generating DI-PEPS.


Next, we discuss another interesting example, which is not constructed from unitary gates, but from the \emph{``plumbing tensor''}~\cite{liu2024topological} which is a special PEPS with $D=\chi^4$.
The local tensor $T_{\mathrm{lbrt}}^{\mathrm{p}}$ is determined by another rank-4 tensor $W$ as 
\begin{equation} \label{eq:plumbing}
T_{\mathrm{lbrt}}^{\mathrm{p}}=T_{\mathrm{lbrt}}^{\mathrm{p}_1\mathrm{p}_2\mathrm{p}_3\mathrm{p}_4}=\delta_{\mathrm{l},\mathrm{p}_1}\delta_{\mathrm{b},\mathrm{p}_2}\delta_{\mathrm{r},\mathrm{p}_3}\delta_{\mathrm{t},\mathrm{p}_4}W_{\mathrm{lbrt}},
\end{equation}
where $\mathrm{p}_i = 1,\dots,\chi$.
Graphically,
$
\begin{tikzpicture}[baseline={([yshift=-0.65ex] current bounding box.center)}, scale=0.8]
\fourPEPSB{0}{0}
\end{tikzpicture}
=
\begin{tikzpicture}[baseline={([yshift=-0.65ex] current bounding box.center)}, scale=0.55]
\draw[very thick](-0.45,-0.9)--(0.45,0.9);
\draw[very thick](-0.9,0)--(0.9,0);
\draw[ thick, fill=myblue, rounded corners=2pt] (-0.4,+0.4) rectangle (+0.4,-0.4);
\Text[x=0,y=0]{\small{$W$}}
\draw[very thick](0.65,0)--(0.65,0.4);
\draw[very thick](-0.65,0)--(-0.65,0.4);
\draw[very thick](0.325,0.65)--(0.325,1.05);
\draw[very thick](-0.4,-0.8)--(-0.4,-0.4);
\end{tikzpicture}
$ 
with 
\begin{tikzpicture}[baseline={([yshift=-0.65ex] current bounding box.center)}, scale=0.5]
\draw[very thick](-0.5,0)--(0.5,0);
\draw[very thick](0,0)--(0,0.5);
\end{tikzpicture} denoting the delta tensor which is non-vanishing only if all three legs take the same value. The elements of the $W$ matrix are restricted by Eqs.~(\ref{eq:ISOTNS}) and (\ref{eq:Dual_ISO_TNS}) (see~\cite{SM} for its specific form). Now, if we place the $W$ tensor at the vertices of a square lattice, the physical degrees of freedom lay on the edges\footnote{Each edge hosts two spins that can be merged into a single one by $\sum_{i=1}^{\chi}\ket{i}\bra{ii}$.}. Notably, the simplest case $\chi=2$ includes the toric code tensor~\cite{kitaev2003fault,liu2024topological} as an example of DI-PEPS. We will return to the topological properties of this class later.

%

\emph{Parameter counting and computational complexity.---}Here we argue about the richness of DI-PEPS, despite them having analytically accessible correlation functions.
As with every PEPS, the representation of the state in terms of tensors is nonunique.
We take into account this so-called gauge freedom by introducing a corresponding canonical form for the family of generalized DI-PEPS (see Eq.~(\ref{eq:GDual_ISO_TNS}) and ~\cite{SM}). The resulting number of free real parameters of a normal\footnote{A PEPS tensor is called normal if it becomes injective after blocking. Generic PEPS are normal~\cite{cirac2021matrix}.}
generalized DI-PEPS (formed by repeating the same tensor) is $2(d-1)\chi^4$.
Compared to the number of free real parameters $2d\chi^4-4\chi^2+2$
of a normal PEPS~\cite{SM}, we see that DI-PEPS cover a large subclass of normal PEPS.

We further consider the computational complexity of DI-PEPS. Although one- and two-point correlators can be efficiently computed, we demonstrate that sampling of a DI-PEPS cannot be efficiently simulated in a classical computer. Following Ref.~\cite{malz2024computational}, we show that the DI-PEPS can encode universal quantum computations with post-selection. Thus, according to Ref.~\cite{Bremner2011}, the output probability distribution cannot be sampled to a multiplicative precision by a classical computer efficiently, unless the polynomial hierarchy collapses to its third level, i.e., $\textsf{postBQP}=\textsf{postBPP}$. 
While the proof and further discussion are available in~\cite{SM}, here we sketch the main points. Based on the previous discussion, we can interpret the PEPS as an 1+1D quantum circuit in the virtual (bond) space, while at the last layer the boundary bond outputs the computation result to physical space. We choose the DI-PEPS constructed from the controlled-dual-unitary gates (example above), thus we encode a dual-unitary circuit into the DI-PEPS. The result follows since the dual-unitary circuits are universal with post-selection~\cite{Suzuki2022computationalpower}. 
This computational complexity also persists for DI-PEPS with further symmetry constraints, such as global $U(1)$ symmetry~\cite{SM}. This is achieved by encoding the above construction into a single block of the symmetric DI-PEPS.



\emph{DI-PEPS, topological states, and locality.---}In this section, we consider in detail a $\chi = 2$ subfamily of the DI-PEPS constructed from the plumbing tensor of Eq.~\eqref{eq:plumbing}, with a special
focus on its topological properties.
To this end, we
impose a $\mathbb Z_{2}$ symmetry on the $W$ tensor 
\begin{equation}
\begin{tikzpicture}[baseline={([yshift=-0.65ex] current bounding box.center)},scale=1.0]
\draw[very thick](-0.5,0)--(0.5,0);
\draw[very thick](-0.25,-0.5)--(0.25,0.5);
\draw[thick, fill=myblue, rounded corners=2pt] (-0.2,0.2) rectangle (0.2,-0.2);
\Text[x=0,y=0]{\small $W$}
\end{tikzpicture}
=
\begin{tikzpicture}[baseline={([yshift=-0.65ex] current bounding box.center)}, scale=1.0]
\draw[very thick](-0.8,0)--(0.8,0);
\draw[very thick](-0.4,-0.8)--(0.4,0.8);
\draw[thick, fill=myblue, rounded corners=2pt] (-0.2,0.2) rectangle (0.2,-0.2);
\draw[thick, fill=myblue] (-0.5,0) circle (0.2);
\draw[thick, fill=myblue] (0.5,0) circle (0.2);
\draw[thick, fill=myblue] (-0.25,-0.5) circle (0.2);
\draw[thick, fill=myblue] (0.25,0.5) circle (0.2);
\Text[x=-0.5,y=0]{\scriptsize{$\sigma^3$}}
\Text[x=0.5,y=0]{\scriptsize{$\sigma^3$}}
\Text[x=-0.25,y=-0.5]{\scriptsize{$\sigma^3$}}
\Text[x=0.25,y=0.5]{\scriptsize{$\sigma^3$}}
\Text[x=0,y=0]{\small $W$}
\end{tikzpicture}
.
\end{equation}
This symmetry restricts its form to 
\begin{equation}
\!W_{\mathrm{lb,rt}}\!\!=\!\!\begin{pmatrix}\sqrt{\alpha} &  &  & \sqrt{1-\alpha}\\
 & \sqrt{\beta} & \sqrt{1-\beta}\\
 & \sqrt{1-\alpha} & \sqrt{\alpha}\\
\sqrt{1-\beta} &  &  & \sqrt{\beta}
\end{pmatrix},\label{eq:plumbing_example_topo}
\end{equation}
with $\alpha,\beta\in[0,1]$, where we omit the complex phases of each element since they are irrelevant
to the topological degeneracy.

Our aim is to probe the topological order of the above subfamily. Each PEPS corresponds to a ground state of a local parent Hamiltonian~\cite{cirac2021matrix}. A PEPS is said to be topologically ordered if the parent Hamiltonian has topological degeneracy in its ground state subspace, which can be characterized by the transfer operator $\mathbb{T}$~\cite{schuch2010peps,PhysRevLett.111.090501}. Parallel to Ref.~\cite{PhysRevLett.111.090501}, here we put the PEPS on a cylinder. We cut out one loop around the cylinder and contract it with its complex conjugate over the physical degrees to form the transfer operator in the doubled virtual space,
\begin{equation}
\mathbb{T}
=
\begin{tikzpicture}[baseline=(current  bounding  box.center), scale=0.7]
\def\lineLength{0.6cm}
\draw[dotted, thick](0,0) ellipse (0.3cm and 1cm);
\draw[very thick](80:0.3cm and 1cm) arc (80:280:0.3cm and 1cm);
\draw[very thick](120:0.3cm and 1cm)--(120:0.5cm and 1.35cm);
\draw[thick, fill=white] (120:0.5cm and 1.35cm) circle (0.08cm); 
\draw[very thick](120:0.3cm and 1cm)--(120:0.5cm and 1.35cm);
\draw[thick, fill=white] (120:0.5cm and 1.35cm) circle (0.08cm);
\draw[very thick](170:0.3cm and 1cm)--(170:0.6cm and 1.35cm);
\draw[thick, fill=white] (170:0.6cm and 1.35cm) circle (0.08cm);
\draw[very thick](220:0.3cm and 1cm)--(220:0.6cm and 1.4cm);
\draw[thick, fill=white] (220:0.6cm and 1.4cm) circle (0.08cm);
\draw (120:0.3cm and 1cm) ++(-\lineLength/2,0) -- ++(\lineLength,0);
\draw (170:0.3cm and 1cm) ++(-\lineLength/2,0) -- ++(\lineLength,0);
\draw (220:0.3cm and 1cm) ++(-\lineLength/2,0) -- ++(\lineLength,0);
\draw[thick,fill=mygreen] (120:0.3cm and 1cm) circle (0.15);
\draw[thick,fill=mygreen] (170:0.3cm and 1cm) circle (0.15);
\draw[thick,fill=mygreen] (220:0.3cm and 1cm) circle (0.15);
\end{tikzpicture}
\;.\label{eq:transfer_OP}
\end{equation}
According to the $\mathbb Z_{2}$ symmetry, the transfer operator can be block diagonalized
into $\mathbb{T}_{p,\phi}^{p',\phi'}$, where $p,p'$ denote the parity of
the ket/bra state, respectively; $\phi,\phi' = 0,\pi$ denote whether a
flux is threading the cylinder, i.e., if an additional $\sigma^3$ is present in the ket/bra virtual space.
Due to the delta tensor in Eq.~\eqref{eq:plumbing},
we immediately see that $p=p'$ and the transformation $\phi(\phi')\to\phi(\phi')+\pi$
does not change the transfer operator. Thus, we only need to distinguish four transfer operators $\mathbb{T}_{\mathrm{e}\phi}^{\mathrm{e}\phi},\mathbb{T}_{\mathrm{o}\phi}^{\mathrm{o}\phi},\mathbb{T}_{\mathrm{e}\phi}^{\mathrm{e}\phi+\pi},\mathbb{T}_{\mathrm{o}\phi}^{\mathrm{o}\phi+\pi}$. 

The key point is that the leading eigenvalues $\lambda_{p\phi}^{p\phi'}$ and the corresponding degeneracy
in each block determine the topological properties of the tensor. This is since they encode, in the thermodynamic limit, the inner product of the states corresponding to different choices of $p,\phi$, which indicates anyon condensation and confinement~\cite{haegeman2015shadows}.

If $\alpha=\beta=\frac{1}{2}$, the resulting state is the toric code~\cite{kitaev2003fault,liu2024topological} with nontrivial topological order. At
this point, $|\lambda_{\mathrm{e}\phi}^{\mathrm{e}\phi}|=|\lambda_{\mathrm{o}\phi}^{\mathrm{o}\phi}|=1,|\lambda_{\mathrm{e}\phi}^{\mathrm{e}\phi+\pi}|=|\lambda_{\mathrm{o}\phi}^{\mathrm{o}\phi+\pi}|=0$
and the blocks with the largest eigenvalues, $\mathbb{T}_{\mathrm{e}\phi}^{\mathrm{e}\phi},\mathbb{T}_{\mathrm{o}\phi}^{\mathrm{o}\phi}$ are non-degenerate. The lines $\alpha=1$ or $\beta=1$,
and the point $\alpha=\beta=0$ are in the trivial phase, since they correspond
to decoupled 1D chains, each in a GHZ state. Those points have exponentially many degeneracies in the leading eigenvalues of the transfer operators $\mathbb{T}_{\mathrm{e}\phi}^{\mathrm{e}\phi},\mathbb{T}_{\mathrm{o}\phi}^{\mathrm{o}\phi}$ with respect to the vertical length $M$ and we refer to them as the GHZ-points.  
In the Supplemental Material~\cite{SM}, we moreover analytically show that except for the GHZ-points mentioned above, it always holds that $|\lambda_{\mathrm{e}\phi}^{\mathrm{e}\phi}|=|\lambda_{\mathrm{o}\phi}^{\mathrm{o}\phi}|>|\lambda_{\mathrm{e}\phi}^{\mathrm{e}\phi+\pi}|=|\lambda_{\mathrm{o}\phi}^{\mathrm{o}\phi+\pi}|$ and the blocks with the largest eigenvalues, $\mathbb{T}_{\mathrm{e}\phi}^{\mathrm{e}\phi},\mathbb{T}_{\mathrm{o}\phi}^{\mathrm{o}\phi}$ are non-degenerate. We achieve that by mapping it to a frustration-free Hamiltonian. We also calculate the full spectrum of the transfer operators along the line $\alpha=\beta$ or $\alpha=1-\beta$ in~\cite{SM}.
Thus we can conclude that the DI-PEPS of Eq.~\eqref{eq:plumbing_example_topo} exhibit the same topological phase as the toric code, except from the GHZ-points. At those points we have a crossing from a topological phase with degeneracy $4$ to a decoupled phase with exponentially many degeneracies. 
This explicitly demonstrates that DI-PEPS contain topological phases.

Another interesting question is the locality of parent Hamiltonians for a given family of PEPS. Note that even states with long-range topological order~\cite{kitaev2003fault,levin2005string,gu2009tensor} admit local parent Hamiltonians.
In principle, the parent Hamiltonian can be obtained by blocking~\cite{cirac2021matrix}, which however may lead to terms with larger supports.
For example, if $d=\chi=2$, generally it is sufficient to block $4\times5$ tensors and the resulting parent Hamiltonian is locally supported on these $20$ sites, which can be slightly decreased by a costly numerical procedure~\cite{PhysRevB.106.035109}. 
Our examples from Eq.~(\ref{eq:plumbing_example_topo}) with non-vanishing $\alpha,\beta$ admit an analytical form of the parent Hamiltonians which are at most $8$-local. In particular, if $\alpha=1-\beta$, this can be improved to a $4$-local Hamiltonian. We illustrate the latter here while the general case follows a similar argument~\cite{SM}. The state with $\alpha=1-\beta$ can be obtained by acting with a single-body operator $\mathcal{U}_v=e^{s\sigma^3}$ on each vertical edge of the toric code, with $s=\frac{1}{4}\ln{\frac{\alpha}{1-\alpha}}$. 
Thus, the parent Hamiltonian of our state can be related to the toric code one via the transformation
$h=\prod_{v}\mathcal{U}_v^{-1\dagger} h_{\mathrm{TC}}\mathcal{U}_v^{-1}$. Here $h$ and $h_\mathrm{TC}$ are local terms of the parent Hamiltonians of our state and the toric code, respectively, and
$v$ belongs to the vertical edges which overlap with $h_{\mathrm{TC}}$. The above transformation does not change the locality of $h$ which remains $4$ as for $h_\mathrm{TC}$~\cite{kitaev2003fault}.

\emph{Discussion and outlook.---}
In this letter, we proposed a new class of PEPS called DI-PEPS, with two isometric conditions. These PEPS allow for efficient computations of local and certain two-point correlation functions. Furthermore, we have shown that this class exhibits interesting physical properties. It has a large number of free parameters, encodes universal quantum computation after post-selection and can represent interesting transitions from topological to trivial order.

It remains an important open question how well the DI-PEPS perform as an ansatz for variational methods, both for ground states and the dynamics. On one hand, the restriction from PEPS to DI-PEPS decreases expressivity. On the other hand the efficient way to obtain reduced density matrices allows for bigger bond dimensions. 

Let us now discuss possible generalizations.
We imposed isometric conditions from top-right and top-left. Actually, one can rotate the tensor and consider the isometric conditions from any two adjacent corners, e.g. top-right and bottom-right. As with iso-PEPS~\cite{PhysRevLett.124.037201}, different choices can be made in different parts of the state.  
However, if one considers the isometric conditions from two diagonal corners (e.g. top-right and bottom-left), the expectation values of local observables are simplified to a contraction of a single tensor. Thus, we expect this class to be less expressive.

Moreover, we can impose additional isometric conditions, from bottom-left and (or) bottom-right.
In this case, higher-point correlation functions are also tractable. One can figure out that for a PEPS with $n$-isometric conditions, all the $(n-1)$-point correlation functions are solvable. 
Although in our study we focus on the 2D square lattice, the generalization to higher dimensions or different lattices is also interesting. We may consider the 3D simple cubic lattice, or the triangle lattice where one can define up to six isometric conditions.

Lastly, the equivalence between PEPS and 1+1D quantum completely-positive maps may guide us to construct new classes of solvable local quantum evolutions for any solvable PEPS. 
As an example, we look at Sequentially Generated States (SGS)~\cite{PhysRevA.77.052306}, which satisfy
$\begin{tikzpicture}[baseline={([yshift=-0.65ex] current bounding box.center)}, scale=0.550]
\fourPEPSG{0}{0}
\MYcircle{0}{0.5}
\MYcircle{0.5}{0}
\end{tikzpicture}
=
\begin{tikzpicture}[baseline={([yshift=-0.65ex] current bounding box.center)}, scale=0.550]
\fourPEPSG{0}{0}
\MYcircle{0}{0.5}
\MYcircle{0.5}{0}
\MYcircle{0}{0.5}
\MYcircle{-0.5}{0}
\draw[very thick](-1.3,0)--(-0.8,0);
\MYcircle{-0.8}{0}
\end{tikzpicture}\;.$
We can use our proposed duality to define a new kind of solvable quantum maps with the property
$\begin{tikzpicture}[baseline=(current  bounding  box.center), scale=0.450]
\Wgategreen{0}{0}
\MYcircle{0.5}{-0.5}
\end{tikzpicture}
=
\begin{tikzpicture}[baseline=(current  bounding  box.center), scale=0.550]
\Wgategreen{0}{0}
\MYcircle{0.5}{-0.5}
\MYcircle{-0.5}{0.5}
\draw[very thick](-0.95,0.95)--(-0.7,0.7);
\MYcircle{-0.7}{0.7}
\end{tikzpicture}.$
This class allows the efficient calculation of local and two-point expectation values (see also ~\cite{wang2024exact}).
Importantly, this class is solvable even for non-trace preserving dynamics, for example including measurements.

\emph{Acknowledgements.---}The authors thank Giacomo Giudice, Yu-Jie Liu, Daniel Malz, Frank Pollmann, Balázs Pozsgay, and Rahul Trivedi for fruitful discussions.
The research is part of the Munich Quantum Valley, which is supported by the Bavarian state government with funds from the Hightech Agenda Bayern Plus. The authors acknowledge funding from the projects FermiQP of the Bildungsministerium für Bildung und Forschung (BMBF).
P.~K. acknowledges financial support from the Alexander von
Humboldt Foundation.

\bibliography{MyCollection}

\clearpage
\onecolumngrid

\appendix
\newcounter{equationSM}
\newcounter{figureSM}
\newcounter{tableSM}
\stepcounter{equationSM}
\setcounter{equation}{0}
\setcounter{figure}{0}
\setcounter{table}{0}
\makeatletter
\renewcommand{\theequation}{\textsc{sm}-\arabic{equation}}
\renewcommand{\thefigure}{\textsc{sm}-\arabic{figure}}
\renewcommand{\thetable}{\textsc{sm}-\arabic{table}}

\begin{center}
{\large{\bf Supplemental Material for\\
 ``Dual-isometric Projected Entangled Pair States''}}
\end{center}

Here we report some useful information complementing the main text. In particular
\begin{itemize}
\item[-] In Section~\ref{app:complexity} we show that with post-selection DI-PEPS can simulate universal quantum computations;
\item[-] In Section~\ref{Secnewadded} we discuss the $U(1)$-symmetric DI-PEPS and its computational complexity;
\item[-] In Section~\ref{app:dimension} we discuss the dimension of the DI-PEPS manifold, generalized DI-PEPS, and canonical form;
\item[-] In Section~\ref{sec:SGS} we show that sequentially generated states are included in generalized DI-PEPS;
\item[-] In section~\ref{app:analyitic} we provide an analytic expression of the transfer operator;
\item[-] In section~\ref{subsec:PH} we discuss an explicit form of parent Hamiltonian;
\item[-] In section~\ref{app:Wtensor} we provide an explicit form of $W$ tensor.
\end{itemize}

\setcounter{equation}{0}
\setcounter{figure}{0}
\setcounter{table}{0}
\makeatletter
\renewcommand{\theequation}{S\arabic{equation}}
\renewcommand{\thefigure}{S\arabic{figure}}


\section{Computational complexity}
\label{app:complexity}
In this section, we investigate the computational complexity of DI-PEPS. We will demonstrate that, with post-selection, DI-PEPS can simulate universal quantum computation. Thus the output probability distribution cannot be sampled to a multiplicative precision by a classical computer efficiently unless the Polynomial Hierarchy collapses to its third level, i.e., $\textsf{postBQP}=\textsf{postBPP}$~\cite{Bremner2011}. Ref.~\cite{malz2024computational} has already shown that using special injective (from bond to physical) PEPS tensors with post-selection can encode universal quantum computations. Their example is in fact a subfamily of DI-PEPS. However, their result requires post-selection on all the 2D lattice sites. Here we make an alternative more explicit construction, which has a further benefit to require post-selection only on a 1D sub-manifold. We achieve this by encoding dual-unitary quantum circuits into the DI-PEPS and referring to known results on their complexity~\cite{Suzuki2022computationalpower}.

We consider the following DI-PEPS with the same boundary condition as in Fig.~(\ref{fig1}b) of the main text
\begin{equation}
\begin{tikzpicture}[baseline=(current  bounding  box.center), scale=0.5]
\leftdownPEPSGR{0}{0}
\leftupPEPSGR{2.5}{5}
\rightupPEPSGR{7.5}{5}
\rightdownPEPSGR{5}{0}
\foreach \i in {0,1}
{
\downPEPSGR{\i+1}{0}
\leftPEPSGR{0.5+\i*0.5}{\i+1}
\upPEPSGR{6.5-\i}{5}
\rightPEPSGR{7-\i*0.5}{4-\i}
}
\fourPEPSGR{1.5}{1}
\fourPEPSGR{6}{4}
\leftPEPSLG{2}{4}
\leftPEPSDB{1.5}{3}
\downPEPSDB{3}{0}
\downPEPSLG{4}{0}
\rightPEPSLG{5.5}{1}
\rightPEPSO{6}{2}
\upPEPSGLG{3.5}{5}
\upPEPSGO{4.5}{5}
\fourPEPSDB{2.5}{1}
\fourPEPSDB{2}{2}
\foreach \i in {0,1,2,3}
{
\fourPEPSO{2.5+\i}{3}
}
\foreach \i in {0,1,2}
{
\fourPEPSO{3+\i}{4}
\fourPEPSO{3+\i}{2}
}
\foreach \i in {0,1}
{
\fourPEPSO{3.5+\i}{1}
}
\end{tikzpicture}.
\end{equation}
Here we use four different kinds of local tensors, denoted by their colors. From the explicit construction shown below, one can verify that all of them satisfy the condition of DI-PEPS. We choose physical dimension $d=16$ and virtual dimension $\chi=2$. However, locally the tensor can have an effective physical dimension\footnote{I.e., the dimension of the image of the tensor, when viewed as a map from virtual to physical space.} lower than $16$, denoted as $d_e$, as only a subspace of the larger physical Hilbert space is relevant.

\emph{Orange tensor:} It is a special case of the class ``controlled-dual unitaries'' proposed
in the main text, but with $d_e=1$, i.e., 
\begin{equation}
T_{\mathrm{lbrt}}^{\mathrm{p}}=\delta_{\mathrm{p},1}V_{\mathrm{tr},\mathrm{lb}}.
\end{equation} Here $V$ is dual unitary~\cite{bertini2019exact}. This tensor can be interpreted as a 1+1D quantum circuit composed of dual-unitary gates acting over the virtual space from left-bottom to right-top. This tensor is the main ingredient for providing the computational complexity of DI-PEPS.

\emph{Light-green tensor:} This tensor with $d_e=16$ is composed of a bipartite physical space,
each with dimension $4$, such that 
\begin{equation}
T_{\mathrm{lbrt}}^{\mathrm{p}_1\mathrm{p}_2}=\braket{\mathrm{lt}|\phi^{p_1}}\braket{\mathrm{rb}|\phi^{p_2}}.
\end{equation}
Here $\ket{\phi^{p_{1(2)}}}$ with $ p_{1(2)} \in \{1,2,3,4\}$ are the four Bell basis states
$\ket{\phi^1}=\frac{\ket{00}+\ket{11}}{\sqrt{2}}$, $\ket{\phi^2}=\frac{\ket{00}-\ket{11}}{\sqrt{2}}$, $\ket{\phi^3}=\frac{\ket{01}+\ket{10}}{\sqrt{2}}$, $\ket{\phi^4}=\frac{\ket{01}-\ket{10}}{\sqrt{2}}$. The state $\ket{\phi^1}$ is also called an EPR pair. The light-green tensor serves as the boundary condition of the dual-unitary circuit. For those on the left-top, we post-select the measurement outcome to $p_2=1$ and on the right-bottom, we post-select the output to $p_1=1$ such that the boundary condition for the dual-unitary circuit is
\begin{equation}
\begin{tikzpicture}[baseline=(current  bounding  box.center), scale=0.5]
\draw[very thick](0,2)--(-1,1);
\draw[very thick](0,0)--(-1,1);
\draw[very thick](0,0)--(-1,-1);
\draw[very thick](-1,-1)--(-0.5,-1.5);
\draw[very thick](0,2)--(-1,3);
\draw[very thick](-1,3)--(-0.5,3.5);
\Wgateblue{0}{0}
\Wgateblue{1}{1}
\Wgateblue{0}{2}
\end{tikzpicture}.
\end{equation}
where the blue rectangles are some dual-unitary gates, following the same notation as in~\cite{bertini2019exact}.
Thus, the dynamic in the virtual space is driven by a dual-unitary quantum circuit.

\emph{Dark-blue tensor:} This tensor with $d_e=4$ is defined as
\begin{equation}
T_{\mathrm{lbrt}}^{\mathrm{p}}=U^\mathrm{p}_\mathrm{lb}\braket{\mathrm{rt}|\phi^1}.
\end{equation}
Here $U$ is an isometry from the left-bottom virtual spaces to physical space, and $\ket{\phi^1}$ corresponds to an EPR pair of the right-top virtual space of the dark-blue tensor. The dark-blue tensors together prepare a chain of EPR pairs that serve as the initial state for the dual-unitary circuit dynamics in the virtual space. The dark-blue tensors also effectively decouple the left-bottom space from the right-top space.

\emph{Grey tensor:} This tensor with $d_e=1$ is defined as 
\begin{equation}
T_{\mathrm{lbrt}}^{\mathrm{p}}=\delta_{\mathrm{p},1}\delta_{\mathrm{l,r}}\delta_{\mathrm{b,t}}.
\end{equation}
This grey tensor is just the SWAP gate in virtual bonds as can be seen in Eq.~(\ref{eq:PEPS_AS_QC}). Pictorially this can be viewed as 
\begin{equation}
T^{\mathrm{p}}_{\mathrm{lbrt}}
=
\begin{tikzpicture}[baseline=(current  bounding  box.center), scale=1.0]
\draw[thick](-0.5,0)--(-0.1,0)--(-0.1,0.1)--(0.1,0.1)--(0.1,0)--(0.5,0);
\draw[thick](-0.25,-0.5)--(0.25,0.5);
\Text[x=-0.65,y=0]{l}
\Text[x=0.65,y=0]{r}
\Text[x=-0.33,y=-0.65]{b}
\Text[x=0.33,y=0.65]{u}
\Text[x=0,y=0.65]{p}
\draw[thick, fill=black](-0.1,0.4) circle (2pt);
\end{tikzpicture}\label{eq:swap_gates_defi}
.
\end{equation}
The information of state is thus swapped to the boundary which is finally read out. 

Therefore, with the above choices, the resulting DI-PEPS can be interpreted as a 1+1D dual-unitary quantum circuit (orange PEPS) over the virtual space with a chain of EPR pairs as the initial state (dark-blue PEPS). The output of this circuit is transmitted to the top-right boundary, which is further transmitted to the physical space and ``read out'' by the last layer of the PEPS. From Ref.~\cite{Suzuki2022computationalpower}, a dual-unitary quantum circuit with initial state a chain of EPR pairs can prepare the cluster state, which serves as the well-known measurement-based universal quantum computation~\cite{PhysRevA.68.022312}. Thus, DI-PEPS under post-selection can simulate universal quantum computation. According to Ref.~\cite{Bremner2011}, quantum states which can encode postselected quantum computation are unlikely to be classically samplable to multiplicative precision, unless $\textsf{postBQP} = \textsf{postBPP}$.

\section{$U(1)$-symmetric DI-PEPS} \label{Secnewadded}
Many physical Hamiltonians have additional global symmetries, like $U(1)$ symmetry in particle-conserving or magnetization-conserving systems. Encoding symmetries into PEPS is important both from the theoretical and numerical standpoint~\cite{singh2011tensor,bauer2011implementing}. Here, we show how to construct $U(1)$-symmetric DI-PEPS.

In the following, we will show that DI-PEPS can encode $U(1)$ symmetry (charge conservation) while maintaining their computational complexity. We start with a simple but pedagogical example of $U(1)$-symmetric DI-PEPS  of spin-1 particles, which we will modify to obtain the final result. For this we consider the case $\chi=4$ and $d=3$. 
The symmetry operator of this system is $e^{i \theta \sum_j Q^{j}}$ ($\theta \in \mathbb R$) where $Q^j$ is the local charge operator acting on site $j$ (e.g., $m_z^j$ for magnetization in the $z$-direction), which has eigenvalues $\{-1,0,1\}$ and the summation is over all the sites in a square lattice. The local physical basis of each site is chosen to correspond to the eigenstates of $Q$. A symmetric state thus has a fixed magnetization $\sum_j Q^{j}$. 
We take each bond space $\mathcal{H}_i$ for $i\in\mathrm{\{l,b,r,t\}}$ to decompose in a direct sum $\mathcal{H}_i=\mathbb{C}^{2}\oplus\mathbb{C}^2$. This can be interpreted as each bond space having two different charges $s_i=\pm\frac{1}{2}$ (labeling the different subspaces in the direct sum) with two states for each charge. We use the pair $(s_i,\alpha)$ to label the state in bond space where $s_i\in\{\frac{1}{2},-\frac{1}{2}\}$ and $\alpha\in\{1,2\}$.

Following~\cite{singh2011tensor,bauer2011implementing}, the tensor $T$ can be decomposed into block-sparse forms. Here we choose the non-vanishing blocks as
\begin{align}
\begin{tikzpicture}[baseline=(current  bounding  box.center), scale=1]
\fourPEPSBDire{0}{0};
\Text[x=0,y=1.2]{\scriptsize{$Q=1$}}
\Text[x=-1.5,y=0]{\small{$s_\mathrm{l}=\frac{1}{2}$}}
\Text[x=1.7, y=0]{\small{$s_\mathrm{r}=-\frac{1}{2}$}}
\Text[x=1.1,y=1.1]{\small{$s_\mathrm{t}=-\frac{1}{2}$}}
\Text[x=-1.1,y=-1.1]{\small{$s_\mathrm{b}=\frac{1}{2}$}}
\end{tikzpicture};&\qquad&
\begin{tikzpicture}[baseline=(current  bounding  box.center), scale=1.0]
\fourPEPSBDire{0}{0};
\Text[x=0,y=1.2]{\scriptsize{$Q=0$}}
\Text[x=-1.6,y=0]{\small{$s_\mathrm{l}=-\frac{1}{2}$}}
\Text[x=1.7, y=0]{\small{$s_\mathrm{r}=\frac{1}{2}$}}
\Text[x=1.1,y=1.1]{\small{$s_\mathrm{t}=-\frac{1}{2}$}}
\Text[x=-1.1,y=-1.1]{\small{$s_\mathrm{b}=\frac{1}{2}$}}
\end{tikzpicture};\nonumber \\
\begin{tikzpicture}[baseline=(current  bounding  box.center), scale=1.0]
\fourPEPSBDire{0}{0};
\Text[x=0,y=1.2]{\scriptsize{$Q=0$}}
\Text[x=-1.5,y=0]{\small{$s_\mathrm{l}=\frac{1}{2}$}}
\Text[x=1.7, y=0]{\small{$s_\mathrm{r}=-\frac{1}{2}$}}
\Text[x=1.1,y=1.1]{\small{$s_\mathrm{t}=\frac{1}{2}$}}
\Text[x=-1.1,y=-1.1]{\small{$s_\mathrm{b}=-\frac{1}{2}$}}
\end{tikzpicture};&\qquad&
\begin{tikzpicture}[baseline=(current  bounding  box.center), scale=1.0]
\fourPEPSBDire{0}{0};
\Text[x=0,y=1.2]{\scriptsize{$Q=-1$}}
\Text[x=-1.6,y=0]{\small{$s_\mathrm{l}=-\frac{1}{2}$}}
\Text[x=1.7, y=0]{\small{$s_\mathrm{r}=\frac{1}{2}$}}
\Text[x=1.1,y=1.1]{\small{$s_\mathrm{t}=\frac{1}{2}$}}
\Text[x=-1.1,y=-1.1]{\small{$s_\mathrm{b}=-\frac{1}{2}$}}
\end{tikzpicture},
\label{eq:charged_symmetry_patten}
\end{align}
and all other blocks except these four vanish. Each charge $s_i$ in the bond space labels a $2$-dimensional subspace. It will be convenient to assign to each bond a direction for the charge current, denoted by arrows in Eq.~(\ref{eq:charged_symmetry_patten}).
By checking each block separately, one can verify that this tensor satisfies the charge conservation law locally, that is 
\begin{equation}
2Q+\sum s_\mathrm{out}-\sum s_\mathrm{in}=2Q+s_\mathrm{r}+s_\mathrm{t}-s_\mathrm{l}-s_\mathrm{b}=0. \label{eq:local_charge_conservation}
\end{equation}
Thus, our construction satisfies the condition for $U(1)$-symmetric PEPS in~\cite{singh2011tensor}. This follows from Eq.~(\ref{eq:local_charge_conservation}), where we sum over all the tensors and all contributions from $s_\mathrm{in}$ and $s_\mathrm{out}$ in the bulk will cancel with each other. Therefore, when we contract all the virtual bonds to obtain a physical state, the latter will automatically preserve the charge, that is, have a definite charge number whose value, in a finite system, is determined by the boundary~\cite{bauer2011implementing}.

To further impose that the construction is a DI-PEPS, it suffices that each block in Eq.~(\ref{eq:charged_symmetry_patten}) is obtained from a dual-unitary gate, as defined in the main text. To be more specific, we require the following form for each block
\begin{equation}
T^\mathrm{p}_\mathrm{lbrt}=T^Q_{(s_\mathrm{l},\alpha),(s_\mathrm{b},\beta),(s_\mathrm{r},\gamma),(s_\mathrm{t},\omega)}=V_{\omega\gamma,\alpha\beta},
\end{equation}
where $Q,s_\mathrm{l},s_\mathrm{b},s_\mathrm{r},s_\mathrm{t}$ match one of the allowed blocks  from Eq.~\eqref{eq:charged_symmetry_patten} and $V$ is a dual-unitary gate, which can be different for different blocks.
It can be directly checked that the above construction satisfies the conditions Eq.~(\ref{eq:ISOTNS}) and Eq.~(\ref{eq:Dual_ISO_TNS}) of DI-PEPS using the definition of a dual-unitary gate. Thus, we have explicitly constructed a DI-PEPS which preserves $U(1)$ symmetry. Interestingly, this $U(1)$-symmetric DI-PEPS also encodes a dual-unitary circuit evolution in the virtual bond space.

Building on this example, we can show that the sampling problem for a DI-PEPS with $U(1)$ symmetry is as hard as for a general DI-PEPS. That is, it remains difficult to classically sample unless $\textsf{postBQP}=\textsf{postBPP}$. We achieve this by showing that any DI-PEPS can be encoded as a block in a $U(1)$-symmetric DI-PEPS.

The construction is very similar to the above one. Given a general DI-PEPS $\mathcal{T}$ whose local tensors $\mathcal{T}^{(x,y)}$ have physical dimension $d$ and virtual dimension $\chi$, we can encode it into a block of $U(1)$-symmetric DI-PEPS $\mathcal{S}$. We use $(x,y)$ to denote the position of a tensor in the square lattice. In our construction, $\mathcal{S}$ is composed of local tensors $\mathcal{S}^{(x,y)}$ whose local physical dimension is $3d$ and bond dimension is $2\chi$. The local charge $Q$ is chosen to have three different eigenvalues $\{-1,0,1\}$, each associated with a $d$-dimension eigenspace, and the constructed state will have a definite quantum number of the total charge $\sum_j Q^{j}$. The local physical basis is chosen as the eigenstates of $Q$ and thus has the structure $\mathbb{C}^d\oplus\mathbb{C}^d\oplus\mathbb{C}^{d}$. Later we will use the pair $(Q,\Gamma)$ to label the physical state where $Q$ denotes three different subspaces in the direct sum and $\Gamma \in\{1,\cdots,d\}$ denotes different states in each subspace, which may be thought as an additional quantum number. The bond space is still given a direct sum  $\mathcal{H}_i=\mathbb{C}^\chi\oplus\mathbb{C}^\chi$ as before, where the two different subspaces are given different charges $s_i=\pm\frac{1}{2}$ and we use the pair $(s_i,\alpha)$ to label the state in the bond space where $\alpha\in\{1,\cdots,\chi\}$.
The allowed blocks for $\mathcal{S}^{(x,y)}$ are also given by Eq.~\eqref{eq:charged_symmetry_patten} and thus the physical state will preserve the $U(1)$ symmetry with symmetry operator $e^{i \theta \sum_j Q^{j}}$. 
Instead of constructing each block from the dual-unitary gate, we now construct each block from a set of tensors $Y_m$ with physical dimension $d$ and virtual bond dimension $\chi$. Each $Y_m$ satisfies the condition of DI-PEPS and $m$ corresponds to different sites and different blocks. That is,
\begin{equation}
(\mathcal{S}^{(x,y)})^\mathrm{p}_\mathrm{lbrt}=(\mathcal{S}^{(x,y)})^{(Q,\Gamma)}_{(s_\mathrm{l},\alpha),(s_\mathrm{b},\beta),(s_\mathrm{r},\gamma),(s_\mathrm{t},\omega)}=(Y_m)^\Gamma_{\alpha\beta\gamma\omega}, \label{eq:construction_of_T}
\end{equation}
where $Q,s_\mathrm{l},s_\mathrm{b},s_\mathrm{r},s_\mathrm{t}$ are one of the blocks allowed in Eq.~(\ref{eq:charged_symmetry_patten}). 
It can be directly checked that the tensors $\mathcal{S}^{(x,y)}$ satisfy conditions of DI-PEPS and thus $\mathcal{S}$ is a $U(1)$-symmetric DI-PEPS.

To encode a general DI-PEPS $\mathcal{T}$ into such $U(1)$-symmetric DI-PEPS $\mathcal{S}$, we require that 
\begin{enumerate}
    \item If $x+y=$ even, the tensor $Y_m$ in the construction Eq.~\eqref{eq:construction_of_T}  corresponding to the block with charge $Q=1$ is $\mathcal{T}_{(x,y)}$.
    \item If $x+y=$ odd, the tensor $Y_m$ in Eq.~\eqref{eq:construction_of_T} corresponding to the block with charge $Q=-1$ is $\mathcal{T}_{(x,y)}$\,
\end{enumerate}
and no restriction on other blocks. With this construction, if we post-select the state to have $Q=1$ for sites with $x+y=$ even and $Q=-1$ for the sites with $x+y=$ odd, we effectively obtain the original DI-PEPS $\mathcal{T}$. It should be noted that we do not need $\mathcal{T}$ to be translationally invariant in the above construction and thus we can encode any DI-PEPS into a $U(1)$-symmetric DI-PEPS with post-selection. According to the discussion in the previous section, where the former can simulate a universal quantum computation with post-selection, the latter can achieve the same through our construction. Thus, we have proven that the hardness in sampling complexity for DI-PEPS persists in the presence of $U(1)$ symmetry, i.e., this class is not efficiently classically samplable to a multiplicative precision unless $\textsf{postBQP}=\textsf{postBPP}$.
Note that the construction at the boundary works automatically as in Sec. \ref{app:complexity}, since tensors at the boundary after post-selection are fixed to match the boundary of $\mathcal{T}$.

Our construction proves that the DI-PEPS can capture  physically relevant symmetries while the sampling problem for it is still hard. This construction may also be generalized to more complicated symmetry groups.

\section{Dimension of the DI-PEPS manifold, generalized families, and canonical form}
\label{app:dimension}

In this section, we explore some mathematical properties of DI-PEPS and their generalization. We discuss the dimension of the underlying manifold and a canonical form. The number of free parameters here always refers to the real degrees of freedom.

\subsection{Dimension of the DI-PEPS manifold}

We first consider the injective PEPS, which is a generic case for $d\geq\chi^4$. Recall that a PEPS tensor is called injective if the corresponding map from virtual to physical space is injective~\cite{cirac2021matrix}. In order to calculate the dimension of the manifold, we recall that a folded PEPS, after contracting the physical degrees of freedom, can be viewed as a completely positive map, shown in Eq.~(\ref{eq:PEPS_AS_QC}). The number of free parameters of
completely positive maps corresponding to DI-PEPS is then equal to the one of the two-unital channels in Ref.~\cite{Kos2023circuitsofspacetime}. The latter can be obtained by perturbing around the totally depolarizing channel~\cite{Kos2023circuitsofspacetime}. For completeness, we repeat this analysis here.

We denote the (conveniently normalized) vectorized identity operator as
\begin{equation}
|
\begin{tikzpicture}[baseline={(0,0.3)}, scale=0.75]
\draw[very thick] (0.0,0.5) -- (0.5,0.5);
\MYcircle{0.5}{0.5}
\end{tikzpicture}
\rangle=
\begin{cases}
\sum_i\ket{ii} & \text{on physical space} \\
\frac{1}{\sqrt{\chi}}\sum_i\ket{ii} & \text{on virtual space}
\end{cases},
\end{equation}
where $\chi$ is the dimension of the virtual space. With this notation, the decomposition of a contracted PEPS over the doubled space can be schematically expressed as
\begin{equation}
\begin{tikzpicture}[baseline={(0,-0.1)}, scale=0.75]
    \fourPEPSG{0}{0}
    \MYcircle{0}{0.5}
\end{tikzpicture}
=a_{\circ\circ\circ\circ}
\begin{tikzpicture}[baseline={(0,-0.1)}, scale=0.75]
\draw[very thick] (0.25,0)--(0.75,0);
\draw[very thick] (-0.25,0)--(-0.75,0);
\draw[very thick] (0,0.25)--(0,0.75);
\draw[very thick] (0,-0.25)--(0,-0.75);
\MYcircle{0.25}{0}
\MYcircle{-0.25}{0}
\MYcircle{0}{0.25}
\MYcircle{0}{-0.25}
\end{tikzpicture}
+a_{\circ\circ\circ -}
\begin{tikzpicture}[baseline={(0,-0.1)}, scale=0.75]
\draw[very thick] (0.25,0)--(0.75,0);
\draw[very thick] (-0.25,0)--(-0.75,0);
\draw[very thick] (0,0.25)--(0,0.75);
\draw[very thick] (0,-0.25)--(0,-0.75);
\MYcircle{0.25}{0}
\MYcircle{-0.25}{0}
\MYcircle{0}{-0.25}
\end{tikzpicture}
+\cdots+a_{\circ\circ --}
\begin{tikzpicture}[baseline={(0,-0.1)}, scale=0.75]
\draw[very thick] (-0.25,0)--(-0.75,0);
\draw[very thick] (0,-0.25)--(0,-0.75);
\draw[very thick] (0,0.75)--(0,0)--(0.75,0);
\MYcircle{-0.25}{0}
\MYcircle{0}{-0.25}
\end{tikzpicture}
+\cdots+a_{----}
\begin{tikzpicture}[baseline={(0,-0.1)}, scale=0.75]
\draw[very thick] (0.75,0)--(-0.75,0);
\draw[very thick] (0,0.75)--(0,-0.75);
\end{tikzpicture}
\label{Eq:decomposition}.
\end{equation}
Here $
\begin{tikzpicture}[baseline={(0,0.3)}, scale=0.75]
\draw[very thick] (0.0,0.5) -- (0.5,0.5);
\end{tikzpicture}
$
denotes an element in the space orthogonal to $|
\begin{tikzpicture}[baseline={(0,0.3)}, scale=0.75]
\draw[very thick] (0.0,0.5) -- (0.5,0.5);
\MYcircle{0.5}{0.5}
\end{tikzpicture}
\rangle$. This space is thus spanned by (vectorized) traceless Hermitian matrices which have $\chi^2-1$ free parameters.

%
%
%
The subscript of $a$ follows the usual order left, bottom, right, top. Therefore the DI-PEPS condition amounts to $a_{\circ\circ\circ\circ}=1$ and $a_{-\circ\circ\circ}=a_{--\circ\circ}=a_{\circ -\circ\circ}=a_{\circ\circ -\circ}=a_{\circ --\circ}=0$ with no restriction on the other components. Thus the space of quantum channels corresponding to DI-PEPS forms a convex subset in the space of completely positive and trace-preserving maps (i.e., quantum channels) with dimension  $\chi^8-2\chi^4+\chi^2 $.
Here the additional requirement that the corresponding channel should be a completely positive map does not change the above parameter counting~\cite{Kos2023circuitsofspacetime}.

To obtain the number of free parameters of the original PEPS $T_{\mathrm{lbrt}}^{\mathrm{p}}$, we need to consider the freedom in the physical index $\mathrm{p}$, which corresponds to the Kraus index in the corresponding representation of a quantum channel. Recall that two Kraus representations are equivalent if and only if they are related by an isometry~\cite{nielsen2002quantum}. In our case, we can select an (arbitrary) set of Kraus operator, $\widetilde{E}_{\mathrm{tr,lb}}^{\mathrm{i}}$ with $i \in \{1,\cdots,\chi^4\}$, to represent the channel corresponding to the contracted DI-PEPS. The original PEPS is thus expressed as
$T_{\mathrm{l}\mathrm{b}\mathrm{r}\mathrm{t}}^{\mathrm{p}}=\sum_{\mathrm{i}}V_{\mathrm{pi}}\widetilde{E}_{\mathrm{tr,lb}}^{\mathrm{i}}$. Here $V$ satisfies $V^\dagger V=I_{\chi^4}$ with $2d\chi^4-\chi^8$ free parameters. Since here we assume the tensor to be injective, $\widetilde{E}_{\mathrm{tr,lb}}^{\mathrm{i}}$ are linearly independent and thus different isometries $V$ must lead to different tensors $T$. The total number of free parameters of injective DI-PEPS is thus
\begin{equation}
2d\chi^4-\chi^8+\chi^8-2\chi^4+\chi^2=2(d-1)\chi^4+\chi^2. \label{eq:dimension_simple_DI}
\end{equation}

For non-injective tensors, it is possible that different isometries lead to the same tensor, so the above counting does not apply directly. Nevertheless, we have numerically calculated the dimension of the tangent space of DI-PEPS and found that Eq.~(\ref{eq:dimension_simple_DI}) holds for generic (non-injective) PEPS for $d<\chi^4$. A notable exception is the permutation-phase gates of the main text [also in Eq.~\eqref{eq:swap_gates_defi}], which have a larger tangent space dimension.

In fact, the following intuitive argument suggests that Eq.~(\ref{eq:dimension_simple_DI}) serves as a lower bound for the number of free parameters. A general PEPS has $2d\chi^4$ free parameters. The isometric condition Eq.~(\ref{eq:ISOTNS}) imposes $\chi^4$ constraints and the dual isometric condition Eq.~(\ref{eq:Dual_ISO_TNS}) imposes another $\chi^4$ constraints. However, these constraints are not linearly independent since both Eqs.~(\ref{eq:ISOTNS}) and~(\ref{eq:Dual_ISO_TNS}) require that 
$\begin{tikzpicture}[baseline={(0,-0.05)}, scale=0.7]
\fourPEPSG{0}{0}
\MYcircle{0}{0.5}
\MYcircle{0.5}{0}
\MYcircle{-0.5}{0}
\MYcircle{0.25}{0.5}
\end{tikzpicture}
=
\begin{tikzpicture}[baseline={(0,-0.1)}, scale=0.7]
\draw[very thick](0,0)--(0,-0.5);
\MYcircle{0}{0}
\end{tikzpicture},$
which includes $\chi^2$ equations. Therefore the number of linearly constraint equations is at most $2\chi^4-\chi^2$. The free parameters are thus, at least, $2d\chi^4-2\chi^4+\chi^2$.

Note that, so far, we have focused on counting parameters that specify a local tensor and not a physical state. We address this in the following, after first introducing more general isometric conditions.


\subsection{Generalized families and canonical form}

The constraints of Eqs.~(\ref{eq:ISOTNS}) and \eqref{eq:Dual_ISO_TNS} for a PEPS are not the most general ones allowing the efficient calculation of expectation values using our strategy. We define \emph{generalized DI-PEPS} by the tensors satisfying the conditions
\begin{equation}
\sum_{\mathrm{r},\mathrm{t},\mathrm{r',t',p}}T_{\mathrm{l}_{1}\mathrm{b}_{1}\mathrm{r}\mathrm{t}}^{\mathrm{p}}S_{\mathrm{r,r'}}R_{\mathrm{t,t'}}{T_{\mathrm{l}_{2}\mathrm{b}_{2}\mathrm{r}'\mathrm{t}'}^{*\mathrm{p}}}=S_{\mathrm{l}_{1},\mathrm{l}_{2}}R_{\mathrm{b}_{1},\mathrm{b}_{2}},\label{eq:general_solvable1}
\end{equation}
\begin{equation}
\sum_{\mathrm{l},\mathrm{t},\mathrm{l',t',p}}T_{\mathrm{l}\mathrm{b}_{1}\mathrm{r}_{1}\mathrm{t}}^{\mathrm{p}}B_{\mathrm{l,l'}}R_{\mathrm{t,t'}}{T_{\mathrm{l'}\mathrm{b}_{2}\mathrm{r}_{2}\mathrm{t}'}^{*\mathrm{p}}}=B_{\mathrm{r}_{1},\mathrm{r}_{2}}R_{\mathrm{b}_{1},\mathrm{b}_{2}}\label{eq:general_solvable2}.
\end{equation}
Here $S,R,B$ are arbitrary matrices. Graphically, this can be expressed as 
\begin{equation} \label{eq:solvable_PEPS_gen_1}
\begin{tikzpicture}[baseline=(current  bounding  box.center), scale=0.7]
\fourPEPSG{0}{0}
\MYdiamond{0.5}{0}
\MYcircle{0}{0.5}
\MYtriangle{0.25}{0.5}
\end{tikzpicture}
=
\begin{tikzpicture}[baseline=(current  bounding  box.center), scale=0.7]
\draw[very thick](-0.75,0)--(-0.25,0);
\draw[very thick](-0.375,-0.75)--(-0.125,-0.25);
\MYdiamond{-0.25}{0}
\MYtriangle{-0.125}{-0.25}
\end{tikzpicture},
\end{equation}
\begin{equation}\label{eq:solvable_PEPS_gen_2}
\begin{tikzpicture}[baseline=(current  bounding  box.center), scale=0.7]
\fourPEPSG{0}{0}
\MYcircle{0}{0.5}
\MYtriangle{0.25}{0.5}
\MYsquareW{-0.5}{0}
\end{tikzpicture}
=
\begin{tikzpicture}[baseline=(current  bounding  box.center), scale=0.7]
\draw[very thick] (0.25,0)--(0.75,0);
\draw[very thick] (-0.125,-0.25)--(-0.375,-0.75);
\MYtriangle{-0.125}{-0.25}
\MYsquareW{0.25}{0}
\end{tikzpicture}.
\end{equation}
Here we use \begin{tikzpicture}[baseline={([yshift=-0.65ex] current bounding box.center)}, scale=0.7]
\MYdiamond{0}{0}
\end{tikzpicture}, 
\begin{tikzpicture}[baseline={([yshift=-0.65ex] current bounding box.center)}, scale=0.7]
\MYtriangle{0}{0}
\end{tikzpicture}, 
\begin{tikzpicture}[baseline={([yshift=-0.65ex] current bounding box.center)}, scale=0.7]
\MYsquareW{0}{0}
\end{tikzpicture}
to represent the vectorized $S,R,B$ matrices, respectively.

We choose open boundary conditions described by rank-$4$ tensors (rank-$3$ at the corners). The requirement for solvability over those boundary tensors will be discussed in the next section. Here we only focus on the bulk tensor, which is assumed to be translationally invariant for simplicity, and consider its degrees of freedom.

Although two PEPS tensors may be different, the resulting states may be the same; this is known as gauge freedom~\cite{cirac2021matrix}. We now use this freedom to bring the matrices $S,R,B$ into a canonical form. In the following, we restrict our discussion to normal PEPS~\cite{cirac2021matrix}. For this class, the gauge freedom is well-understood~\cite{perez2010characterizing} (including for open boundary conditions~\cite{molnar2018normal}).
In particular, two normal tensors represent the same state if and only if they can be connected by a gauge transformation
\begin{equation}
\begin{tikzpicture}[baseline={([yshift=-0.65ex] current bounding box.center)}, scale=0.8]
\fourPEPSB{0}{0}
\end{tikzpicture}
\to
\begin{tikzpicture}[baseline={([yshift=-0.65ex] current bounding box.center)}, scale=0.8]
\draw[very thick](-1.0,0)--(1.0,0);
\draw[very thick](-0.5,-1.0)--(0.5,1.0);
\fourPEPSB{0}{0}
\draw[ thick, fill=myblue, rounded corners=2pt] (-0.85,+0.25) rectangle (-0.35,-0.25);
\Text[x=-0.6,y=0]{\scriptsize{$Q$}}
\draw[ thick, fill=myblue, rounded corners=2pt] (0.35,+0.25) rectangle (0.85,-0.25);
\Text[x=0.6,y=0]{\scriptsize{$Q^{\text{-}1}$}}
\draw[thick, fill=myblue, rounded corners=2pt] (0.05,0.85) rectangle (0.55,0.35);
\Text[x=0.3,y=0.6]{\scriptsize{$J$}}
\draw[thick, fill=myblue, rounded corners=2pt] (-0.05,-0.85) rectangle (-0.55,-0.35);
\Text[x=-0.3,y=-0.6]{\scriptsize{$J^{\text{-}1}$}}
\end{tikzpicture},
\end{equation}
(see Ref.~\cite{molnar2018normal}), with $Q,J$ arbitrary invertible matrices. After the gauge transformation, the corresponding eigenvectors $S,R,B$ in Eqs. (\ref{eq:general_solvable1}) and (\ref{eq:general_solvable2}) will also be changed. In the following, we use this gauge freedom to bring the matrices $S,R,B$ into a canonical form.

A normal PEPS can be blocked into a large rectangle tensor such that it is injective. The blocked region is depicted as
\begin{equation}
\begin{tikzpicture}[baseline=(current  bounding  box.center), scale=0.7]
\fourPEPSB{1}{0}
\fourPEPSB{2}{0}
\Text[x=3,y=0]{$\cdots$}
\fourPEPSB{4}{0}
\Text[x=2,y=-1]{$\vdots$}
\fourPEPSB{0}{-2}
\fourPEPSB{1}{-2}
\Text[x=2,y=-2]{$\cdots$}
\fourPEPSB{3}{-2}
\end{tikzpicture}. \label{eq:blocked_TNS}
\end{equation}
The corresponding folded tensor after contracting the physical degrees then defines a completely positive map, when interpreted from top-right to bottom-left. We recognize that
\begin{equation}
\begin{tikzpicture}[baseline=(current  bounding  box.center), scale=0.7]
\draw[very thick](0,1)--(0,0.5);
\draw[very thick](1,1)--(1,0.5);
\draw[very thick](3,1)--(3,0.5);
\MYtriangle{0}{1}
\MYtriangle{1}{1}
\MYtriangle{3}{1}
\Text[x=2,y=0.75]{$\cdots$}
\draw[very thick](3,0)--(3.5,0);
\draw[very thick](3,-2)--(3.5,-2);
\Text[x=3.25,y=-1]{$\vdots$}
\MYdiamond{3.5}{0}
\MYdiamond{3.5}{-2}
\end{tikzpicture}\label{eq:eigenvector_of_TO_blocked}
\end{equation}
is an eigenvector with the eigenvalue $1$. We assume that this map has no eigenvalue larger than $1$ for a large enough blocking, which will be justified in the next paragraph.
Since this blocked tensor is injective, we can use a lemma from~\cite{PhysRevB.101.094304} which states that the completely positive map corresponding to an injective tensor is primitive and thus its leading eigenvector must be unique and strictly positive definite. This implies that the matrices $S,R$ are strictly positive definite. Parallel to Ref.~\cite{PhysRevB.101.094304}, we choose $J=R^{\frac{1}{2}}$ and $Q=S^{-\frac{1}{2}}$ such that for the new tensor $\widetilde{T}$, Eq.~(\ref{eq:general_solvable1}) reduces to the isometric condition Eq.~(\ref{eq:ISOTNS}) and Eq.~(\ref{eq:general_solvable2}) simplifies to a new one with $R$ replaced by the identity matrix and $B$ replaced by $\widetilde{B}=S^{\frac{1}{2}}BS^{\frac{1}{2}}$.

Here we argue that the assumption that the map after large enough blocking has no eigenvalue larger than $1$ is sensible. To this end, we require without loss of generality that the state is normalized in the thermodynamic limit. Equation~(\ref{eq:blocked_TNS}) is further blocked horizontally and then vertically to form a corner transfer operator~\cite{vanderstraeten2016gradient}, i.e.,
the basic tensor of the corner transfer operator is now replaced by the blocked one of Eq.~(\ref{eq:blocked_TNS}). According to Ref.~\cite{vanderstraeten2016gradient}, the norm of the state can be approximated by the power of the leading eigenvalue of the corner transfer operator, which suggests that no eigenvalue larger than $1$ is allowed as long as the state is normalized in the thermodynamic limit. For the corner transfer operator constructed above, Eq.~(\ref{eq:eigenvector_of_TO_blocked}) with extension to infinity is just the eigenvector with eigenvalue $1$, which means that it is the leading eigenvector. Since the corner transfer operator is formed by folding an injective tensor [constructed from Eq.~(\ref{eq:blocked_TNS})], we immediately obtain that $S,R$ are strictly positive definite~\cite{PhysRevB.101.094304} and everything else follows as before.

Now, the only gauge freedom remaining which does not change Eq.~(\ref{eq:ISOTNS}) is the unitary gauge, i.e., we can choose $J,Q$ to be both unitary. Since the matrix $\widetilde{B}$ is positive definite (following a similar argument), this gauge choice can be used to diagonalize it. For simplicity, we assume that the eigenbasis of $\widetilde{B}$ is non-degenerate, which is the generic case. Thus, the canonical form of the generalized DI-PEPS can be expressed as (here we use the symbol $T$ to also represent the tensor after the gauge transformations from above were implemented)
\begin{equation}
\sum_{\mathrm{r},\mathrm{t},\mathrm{p}}T_{\mathrm{l}_{1}\mathrm{b}_{1}\mathrm{r}\mathrm{t}}^{\mathrm{p}}{T_{\mathrm{l}_{2}\mathrm{b}_{2}\mathrm{r}\mathrm{t}}^{*\mathrm{p}}}=\delta_{\mathrm{l}_{1},\mathrm{l}_{2}}\delta_{\mathrm{b}_{1},\mathrm{b}_{2}},\label{eq:canonical_form}
\end{equation}
\begin{equation}
\sum_{\mathrm{l},\mathrm{t},\mathrm{p}}T_{\mathrm{l}\mathrm{b}_{1}\mathrm{r}_{1}\mathrm{t}}^{\mathrm{p}}\Lambda_{\mathrm{l,l}}{T_{\mathrm{l}\mathrm{b}_{2}\mathrm{r}_{2}\mathrm{t}}^{*\mathrm{p}}}=\Lambda_{\mathrm{r}_1,\mathrm{r}_1}\delta_{\mathrm{r}_{1},\mathrm{r}_{2}}\delta_{\mathrm{b}_{1},\mathrm{b}_{2}} \label{eq:canonical_form2}
\end{equation}
for some diagonal matrix $\Lambda$. Or equivalently in graphical notation
\begin{equation} \label{eq:solvable_PEPS_canonical_1}
\begin{tikzpicture}[baseline=(current  bounding  box.center), scale=0.7]
\fourPEPSG{0}{0}
\MYcircle{0.5}{0}
\MYcircle{0}{0.5}
\MYcircle{0.25}{0.5}
\end{tikzpicture}
=
\begin{tikzpicture}[baseline=(current  bounding  box.center), scale=0.7]
\draw[very thick](-0.75,0)--(-0.25,0);
\draw[very thick](-0.375,-0.75)--(-0.125,-0.25);
\MYcircle{-0.25}{0}
\MYcircle{-0.125}{-0.25}
\end{tikzpicture},
\end{equation}
\begin{equation} \label{eq:solvable_PEPS_canonical_2}
\begin{tikzpicture}[baseline=(current  bounding  box.center), scale=0.7]
\fourPEPSG{0}{0}
\MYcircle{0}{0.5}
\MYcircle{0.25}{0.5}
\MYsquareW{-0.5}{0}
\end{tikzpicture}
=
\begin{tikzpicture}[baseline=(current  bounding  box.center), scale=0.7]
\draw[very thick] (0.25,0)--(0.75,0);
\draw[very thick] (-0.125,-0.25)--(-0.375,-0.75);
\MYcircle{-0.125}{-0.25}
\MYsquareW{0.25}{0}
\end{tikzpicture},
\end{equation}
with the white square now the vectorized form of $\Lambda$. At this stage, the remaining gauge corresponds to $Q$ being the diagonal phase matrix in the same eigenbasis as $\Lambda$,  
and $J$ being any unitary matrix.

The number of free parameters associated with the physical state described by a uniform normal generalized DI-PEPS [i.e., satisfying Eqs.~\eqref{eq:solvable_PEPS_gen_1} and \eqref{eq:solvable_PEPS_gen_2}] can be estimated 
from the free parameters of the canonical form [Eqs.~\eqref{eq:solvable_PEPS_canonical_1} and \eqref{eq:solvable_PEPS_canonical_2}] minus the remaining gauge. Eq.~(\ref{eq:canonical_form2}) or (\ref{eq:solvable_PEPS_canonical_2}) with any fixed diagonal matrix $\Lambda$ allows a similar decomposition as in Eq.~(\ref{Eq:decomposition}).  
We just need to replace the right empty bullet $\begin{tikzpicture}[baseline={([yshift=-0.65ex] current bounding box.center)}, scale=0.7]
\draw[very thick](0,0)--(0.5,0);
\MYcircle{0}{0}
\end{tikzpicture}$ in Eq.~(\ref{Eq:decomposition}) with $\begin{tikzpicture}baseline={([yshift=-0.65ex] current bounding box.center)}, scale=0.7]
\draw[very thick](0,0)--(0.5,0);
\draw[thick, fill=white](0.08,0.08) rectangle (-0.08,-0.08);
\end{tikzpicture}$ (vectorized $\Lambda$),
and also the left solid line in Eq.~(\ref{Eq:decomposition}) with an element in the orthogonal space of $\Lambda$. This corresponds to the biorthogonal decomposition in the horizontal direction where the left and right eigenvectors are different. 
For example, the first term would be
\begin{tikzpicture}[baseline={([yshift=-0.65ex] current bounding box.center)}, scale=0.7]
\draw[very thick] (0.25,0)--(0.75,0);
\draw[very thick] (-0.25,0)--(-0.75,0);
\draw[very thick] (0,0.25)--(0,0.75);
\draw[very thick] (0,-0.25)--(0,-0.75);
\MYsquareW{0.25}{0}
\MYcircle{-0.25}{0}
\MYcircle{0}{0.25}
\MYcircle{0}{-0.25}
\end{tikzpicture}.
Thus, for any fixed $\Lambda$, a generic PEPS satisfying Eqs. (\ref{eq:canonical_form}), (\ref{eq:canonical_form2}) has the same number of free parameters as the one satisfying Eqs.~(\ref{eq:ISOTNS}), (\ref{eq:Dual_ISO_TNS}), but with $\Lambda$ contributing additional $\chi-1$ free parameters (here we have subtracted the global scaling $\Lambda\to c \Lambda$ with $c$ a constant).

Combining all the things together, we arrive at the total number of free parameters associated with the generic physical state (i.e., the one satisfying all previous assumptions) as
\begin{align}
&2(d-1)\chi^4+\chi^2+\chi-1-(\chi-1+\chi^2-1)-1 \nonumber\\
=& 2(d-1)\chi^4.
\end{align}
Here the term in the second parenthesis represents the gauge freedom of $J,Q$ (whose global phase is irrelevant since it cancels with $J^{-1},Q^{-1}$). The last $-1$ comes from the fact that two tensors $T$ differing by a phase represent the same physical state.

For comparison, we also derive the number of free parameters of a physical state associated with a generic (i.e., normal) PEPS. For such a PEPS, the rank-$5$ tensor $T$ has $2d\chi^4$ free parameters. Now, because of the gauge freedom, $J,Q$ can be chosen as arbitrary invertible matrices. Moreover, the total scaling factor of $J,Q$ is irrelevant since it cancels with $J^{-1},Q^{-1}$. Thus, each of $J,Q$ has $2\chi^2-2$ free parameters. A general PEPS is not normalized, which corresponds to subtracting two additional free parameters from $T$. The number of free parameters is thus $2d\chi^4-4\chi^2+2$.

\subsection{Solvable boundary tensors}

Lastly, we consider the solvable boundary tensors associated with Eqs. (\ref{eq:canonical_form}) and (\ref{eq:canonical_form2}). We explicitly analyze the left-boundary tensor; the rest is understood similarly. Here, we assume that the left boundary is described by a rank-$4$ tensor $L^{\mathrm{p}}_{\mathrm{brt}}$
as 
\begin{equation}
\begin{tikzpicture}[baseline=(current  bounding  box.center), scale=0.7]
\threePEPSB{0}{0}
\Text[x=0.65,y=0]{\small $\mathrm{r}$}
\Text[x=-0.33,y=-0.65]{\small $\mathrm{b}$}
\Text[x=0.33,y=0.65]{\small $\mathrm{t}$}
\Text[x=0,y=0.7]{\small $\mathrm{p}$}
\end{tikzpicture}
=
L^{\mathrm{p}}_{\mathrm{brt}}.
\end{equation}
The boundary condition is solvable if $L$ satisfies 
\begin{equation}
\begin{tikzpicture}[baseline=(current  bounding  box.center), scale=0.7]
\threePEPSG{0}{0}
\MYcircle{0}{0.5}
\MYcircle{0.25}{0.5}
\end{tikzpicture}
=
\begin{tikzpicture}[baseline=(current  bounding  box.center), scale=0.7]
\draw[very thick] (0.25,0)--(0.75,0);
\draw[very thick] (-0.125,-0.25)--(-0.375,-0.75);
\MYcircle{-0.125}{-0.25}
\MYsquareW{0.25}{0}
\end{tikzpicture},
\label{eq:solvable_BC}
\end{equation}
where the green ball depicts the doubled tensor $L^{*\mathrm{p}}_{\mathrm{brt}} \otimes L^{{\mathrm{p}}}_{\mathrm{brt}}$.
One can immediately see that the boundary condition shown in Fig.~(\ref{fig1}b) of the main text is a special case of Eq.~(\ref{eq:solvable_BC}) with $\Lambda$ replaced by the identity and $L^{\mathrm{p}}_{\mathrm{brt}}=\delta_{\mathrm{p,r}}\delta_{\mathrm{t,b}}$.

\section{Sequentially Generated States are included in generalized DI-PEPS}
\label{sec:SGS}
In this section, we prove that all sequentially generated states (SGS)~\cite{PhysRevA.77.052306} are actually included in our generalized DI-PEPS class.
A SGS can be expressed as~\cite{PhysRevLett.128.010607}
\begin{equation}
T^\mathrm{p}_\mathrm{lbrt}=
\begin{tikzpicture}[baseline=(current  bounding  box.center), scale=0.7]
\Wgateblue{0}{0}
\Wgateblue{-1}{1}
\Text[x=0.65,y=-0.65]{$\ket{0}$}
\Text[x=0.65,y=0.55]{r}
\Text[x=-0.55,y=-0.65]{l}
\Text[x=-1.55,y=0.35]{b}
\Text[x=-0.35,y=1.55]{t}
\Text[x=-1.55,y=1.55]{p}
\Text[x=0,y=0]{\scriptsize{$V$}}
\Text[x=-1,y=1]{\scriptsize{$U$}}
\end{tikzpicture}
=
\sum_k U^{\mathrm{pt}}_\mathrm{bk}V^\mathrm{kr}_\mathrm{l0}.
\end{equation}
where $U, V$ are both unitary gates. The isometric condition~(\ref{eq:ISOTNS}) can be checked directly. For the other direction, when we contract the physical and top indices, we obtain
\begin{equation}
\sum_{\mathrm{p,t}}T^{\mathrm{p}}_{\mathrm{l}_1\mathrm{b}_1\mathrm{r}_1\mathrm{t}}T^{*\mathrm{p}}_{\mathrm{l}_2\mathrm{b}_2\mathrm{r}_2\mathrm{t}}=\delta_{\mathrm{b}_1\mathrm{b}_2}\sum_\mathrm{k}V^{\mathrm{k}\mathrm{r}_1}_{\mathrm{l}_10}V^{*\mathrm{k}\mathrm{r}_2}_{\mathrm{l}_20}, \label{eq:SGS_dual_condi}
\end{equation}
where we have already used the unitarity of $U$. Since $V$ is also a unitary, the sum in the right hand side of Eq.~(\ref{eq:SGS_dual_condi}) is actually the Stinespring representation of a quantum channel with input indices $l_1,l_2$ and output indices $r_1,r_2$. Therefore, it always has a fixed point with eigenvalue $1$~\cite{wolf2012quantum}, that is
\begin{equation}
\sum_{\mathrm{k,l}_1,\mathrm{l}_2}V_{\mathrm{l}_10}^{\mathrm{kr}_1}\tilde{B}_{\mathrm{l}_1\mathrm{l}_2}V_{\mathrm{l}_20}^{*\mathrm{kr}_2}=\tilde{B}_{\mathrm{r}_1\mathrm{r}_2}. \label{eq:fixed_pt_SRe}
\end{equation}
Combining Eqs.~(\ref{eq:fixed_pt_SRe}) and (\ref{eq:SGS_dual_condi}) together, we immediately obtain that 
\begin{equation}
\sum_{\mathrm{p,t,l}_1,\mathrm{l}_2}T^{\mathrm{p}}_{\mathrm{l}_1\mathrm{b}_1\mathrm{r}_1\mathrm{t}}\tilde{B}_{\mathrm{l}_1\mathrm{l}_2}
T^{*\mathrm{p}}_{\mathrm{l}_2\mathrm{b}_2\mathrm{r}_2\mathrm{t}}=\delta_{\mathrm{b}_1\mathrm{b}_2}\tilde{B}_{\mathrm{r}_1\mathrm{r}_2},
\end{equation}
which is nothing but our definition of the generalized DI-PEPS, see Eqs.~(\ref{eq:general_solvable1}) and (\ref{eq:general_solvable2}) and the discussions below. Thus, we have proved that SGS automatically satisfy the condition of generalized DI-PEPS and is a subclass of the latter.

\section{Analytic expression of the transfer operator}
\label{app:analyitic}

Here we {derive analytic expressions for} the transfer operators of Eq.~(\ref{eq:transfer_OP}).

Due to the presence of the delta tensor in the plumbing tensor Eq.~\eqref{eq:plumbing}, folding results in bipartite spaces, but with each part carrying a copy of the other one in the computational basis. That is, contracting two delta tensors
\begin{equation}
\begin{tikzpicture}[baseline={(0,-0.1)}, scale=0.7]
\draw[very thick](-0.5,0.5)--(0.5,0.5);
\draw[very thick](-0.5,-0.5)--(0.5,-0.5);
\draw[very thick](0,0.5)--(0,-0.5);
\end{tikzpicture}
=
\ket{00}\bra{00}+\ket{11}\bra{11},
\end{equation}
results in a rank-2 projector.
Thus, for convenience we only work in the subspace of its image and view $\ket{00}\to\ket{0}$, $\ket{11}\to\ket{1}.$ Therefore, the green tensor in Eq.~(\ref{eq:transfer_OP}) (PEPS tensor after contracting the physical degrees) corresponds to a different plumbing tensor,
obtained by squaring $W$ element-wise as

\begin{equation}
\!\widetilde{W}_{\mathrm{lb,rt}}\!\!=\!\!\begin{pmatrix}\alpha &  &  & 1-\alpha\\
 & \beta & 1-\beta\\
 & 1-\alpha & \alpha\\
1-\beta &  &  & \beta
\end{pmatrix}.\label{eq:doubled_W_tensor}
\end{equation}
For the latter use, we reorder $\widetilde{W}$ as

\begin{align}
\!\widetilde{W}_{\mathrm{rt,lb}}\!\!=&\!\!\begin{pmatrix}\alpha &  &  & 1-\beta\\
 & \beta & 1-\alpha\\
 & 1-\beta & \alpha\\
1-\alpha &  &  & \beta
\end{pmatrix} \nonumber \\
=& I_H\otimes\begin{pmatrix}
\alpha \\
 & \beta
\end{pmatrix}_V+\sigma^1_H\otimes\begin{pmatrix}
&1-\beta \\
1-\alpha
 \end{pmatrix}_V. \label{eq:doubled_space_product}
\end{align}
Here the last line suggests that we can interpret the tensor $\widetilde{W}$ as a joint operator acting on the product of two $2$-dimensional spaces. We denote these spaces as horizontal, indexed with $H$, and vertical, indexed with $V$. The former corresponds to the indices $\mathrm{l,r}$ of the original $\widetilde{W}$ tensor and the latter corresponds to the $\mathrm{b,t}$ indices. We emphasize that, in the transfer operator given by Eq.~(\ref{eq:transfer_OP}), each tensor $\widetilde{W}$ at different positions $i$ has its own horizontal space. But all tensors share a common vertical space since they are connected vertically. The transfer operator with $\phi'=\phi$ thus can be expressed as 
\begin{equation}
\mathbb{T}_\phi^\phi=\mathrm{Tr}_V(\prod_i^M \widetilde{W}_i) \label{eq:new_express_T}
\end{equation}
with $M$ the number of sites in the transfer operator. Without loss of generality, we assume $M$ to be even. Now the transfer operator acts on a product of $M$ horizontal spaces.
Substituting Eq.~(\ref{eq:doubled_space_product}) into Eq.~(\ref{eq:new_express_T}) we directly see that $\mathbb{T}$ only consists of Identity and $\sigma^1$ operators. Moreover, due to the trace in Eq.~(\ref{eq:new_express_T}), $\mathbb{T}$ must include an even number of $\sigma^1$. After some algebraic calculations, we can obtain the expression
\begin{align}
\mathbb{T}_\phi^\phi=&I(\alpha^M+\beta^M)+\sum_{1\leq i<j\leq M}\sigma^1_i\sigma^1_j(1-\alpha)(1-\beta)(\beta^{j-i-1}\alpha^{M+i-j-1}+\alpha^{j-i-1}\beta^{M+i-j-1})\nonumber \\
+&\sum_{1\leq i<j<k<n\leq M}\sigma^1_i\sigma^1_j\sigma^1_k\sigma^1_n(1-\alpha)^2(1-\beta)^2(\beta^{j-i+n-k-2}\alpha^{M+i-n+k-j-2}+\alpha^{j-i+n-k-2}\beta^{M+i-n+k-j-2})+\cdots\nonumber\\
+2&\prod_{i=1}^M\sigma^1_i(1-\alpha)^{M/2}(1-\beta)^{M/2} \label{eq:general_expression_trans}
\end{align}
where $\cdots$ denotes terms with a larger (even) number of $\sigma^1$ matrices.
Since Eq.~(\ref{eq:general_expression_trans}) is frustration-free and each factor is non-negative, we immediately see that there are two degenerate leading eigenvectors $\ket{\psi_{+}}=\otimes_i^M\ket{+}_i,\ket{\psi_{-}}=\otimes_i^M\ket{-}_i$. Here $\ket{+}=\frac{\ket{0}+\ket{1}}{\sqrt{2}},\ket{-}=\frac{\ket{0}-\ket{1}}{\sqrt{2}}$ are the eigenvectors of $\sigma^1$. The linear combinations of these two vectors $\frac{\ket{\psi_{+}}+\ket{\psi_{-}}}{\sqrt{2}}$, $\frac{\ket{\psi_{+}}-\ket{\psi_{-}}}{\sqrt{2}}$ have even and odd parity, corresponding to $\mathbb{T}_{\mathrm{e}\phi}^{\mathrm{e}\phi},\mathbb{T}_{\mathrm{o}\phi}^{\mathrm{o}\phi}$, respectively. Thus, $\lambda_{\mathrm{e}\phi}^{\mathrm{e}\phi}=\lambda_{\mathrm{o}\phi}^{\mathrm{o}\phi}$ always holds.

As long as we are not at the GHZ-points, where $\alpha=1$ or $\beta=1$ or $\alpha=\beta=0$, each coefficient in Eq.~(\ref{eq:general_expression_trans}) is strictly positive. Any other state not spanned by $\ket{\psi_{+}}$ and $\ket{\psi_{-}}$ must violate some terms in Eq.~(\ref{eq:general_expression_trans}) and thus decrease the leading eigenvalue. 
Therefore, the leading eigenvalue of $\mathbb{T}$ is at most $2$-fold degenerate, each corresponding to one of the blocks $\mathbb{T}_{\mathrm{e}\phi}^{\mathrm{e}\phi},\mathbb{T}_{\mathrm{o}\phi}^{\mathrm{o}\phi}$ separately. 

If $\phi'=\phi+\pi$, the transfer operator is given by
\begin{equation}
\mathbb{T}_{\phi}^{\phi+\pi}=\mathrm{Tr}_V(\prod_i^M\widetilde{W}_i\sigma^3_V).
\end{equation}
After reasoning as above, the expression is 
\begin{align}
\mathbb{T}_\phi^{\phi+\pi}=&I(\alpha^M-\beta^M)+\sum_{1\leq i<j\leq M}\sigma^1_i\sigma^1_j(1-\alpha)(1-\beta)(\beta^{j-i-1}\alpha^{M+i-j-1}-\alpha^{j-i-1}\beta^{M+i-j-1})\nonumber \\
+&\sum_{1\leq i<j<k<n\leq M}\sigma^1_i\sigma^1_j\sigma^1_k\sigma^1_n(1-\alpha)^2(1-\beta)^2(\beta^{j-i+n-k-2}\alpha^{M+i-n+k-j-2}-\alpha^{j-i+n-k-2}\beta^{M+i-n+k-j-2})+\cdots, \label{eq:general_TO_with_pi}
\end{align}
which has a similar form as Eq.~(\ref{eq:general_expression_trans}) but with some coefficients changed from positive to negative. Thus, as long as we are not at the GHZ-points, the leading eigenvalue of Eq.~(\ref{eq:general_TO_with_pi}) must be lower than that of Eq.~(\ref{eq:general_expression_trans}). Since it still has the symmetry $\otimes_i^M\sigma^3_i$, each eigenvalue is at least twice degenerate, one corresponds to $\mathbb{T}_{\mathrm{e}\phi}^{\mathrm{e}\phi+\pi}$ and the other to $\mathbb{T}_{\mathrm{o}\phi}^{\mathrm{o}\phi+\pi}$. Thus, we proved that as long as we are not at the GHZ-points, $|\lambda_{\mathrm{e}\phi}^{\mathrm{e}\phi}|=|\lambda_{\mathrm{o}\phi}^{\mathrm{o}\phi}|>|\lambda_{\mathrm{e}\phi}^{\mathrm{e}\phi+\pi}|=|\lambda_{\mathrm{o}\phi}^{\mathrm{o}\phi+\phi}|$ and the blocks with the largest eigenvalues $\mathbb{T}_{\mathrm{e}\phi}^{\mathrm{e}\phi},\mathbb{T}_{\mathrm{o}\phi}^{\mathrm{o}\phi}$ are nondegenerate. This implies that those states are in the same topological order as the toric code.

We illustrate the above ideas with two examples. 
The first example is $\beta=1-\alpha$. In this case, with a straightforward calculation, Eq.~(\ref{eq:general_expression_trans}) can be rewritten as 
\begin{equation}
\mathbb{T}_\phi^\phi=\prod_i^{M-1}\left(\alpha I+ (1-\alpha)\sigma^1_i\sigma^1_{i+1}\right)\left(\alpha I+(1-\alpha)\sigma^1_M\sigma^1_1\right).
\end{equation}
This transfer matrix has the leading eigenvalue $1$ for eigenvectors $\ket{\psi_{+}}, \ket{\psi_{-}}$ as the only leading eigenvectors. The second leading eigenvalue is $(1-2\alpha)^2$ with the eigenvectors obtained by replacing one $\ket{+}$ ($\ket{-}$) from $\ket{\psi_{+}}$ ($\ket{\psi_{-}}$) with $\ket{-}$ ($\ket{+}$). As $\alpha\to0$, the gap closes as $ 4\alpha$. Similarly, Eq.~(\ref{eq:general_TO_with_pi}) simplifies to 
\begin{equation}
\mathbb{T}_\phi^{\phi+\pi}=\prod_i^{M-1}\left(\alpha I+ (1-\alpha)\sigma^1_i\sigma^1_{i+1}\right)\left(\alpha I-(1-\alpha)\sigma^1_M\sigma^1_1\right).
\end{equation}
The absolute value of its leading eigenvalue is $|1-2\alpha|$ which is always smaller than $1$ as long as $\alpha \neq 0,1$.

The second example is $\beta=\alpha$ where, after straightforward calculations, Eq.~(\ref{eq:general_expression_trans}) can be rewritten as 
\begin{equation}
\mathbb{T}_\phi^\phi=\prod_i^M\left(\alpha I+(1-\alpha)\sigma^1_i\right)+\prod_i^M\left(\alpha I-(1-\alpha)\sigma^1_i\right).
\end{equation}
In the above equation, the sum of the two terms ensures the symmetry $\otimes_i^M\sigma^3_i$, thus only terms with an even number of $\sigma^1$ can survive in the final result. This transfer operator has the leading eigenvalue $1+(2\alpha-1)^M$. The absolute value of the second one is $|2\alpha-1|+|2\alpha-1|^{M-1}$. The corresponding eigenvectors are the same as in the first example. As long as $\alpha\neq 0,1$, the gap does not close. When $\alpha\to 0$, the gap closes as $4(M-1)\alpha^2$. On the other hand, Eq.~(\ref{eq:general_TO_with_pi}) vanishes 
\begin{equation}
\mathbb{T}_\phi^{\phi+\pi}=0.
\end{equation}

\section{Parent Hamiltonian}\label{subsec:PH}

In this section, we discuss the parent Hamiltonian of the DI-PEPS constructed from the plumbing tensor Eq.~(\ref{eq:plumbing_example_topo}) with $\alpha,\beta \in (0,1)$.  The special point $\alpha=\beta=1/2$  corresponds to the toric code $\ket{\mathrm{TC}}$, which has the parent Hamiltonian 
\begin{equation}
H=\sum_v \mathcal{A}_v+\sum_p \mathcal{B}_p,
\end{equation}
where $\mathcal{A}_v=I-\prod_{i\in v}\sigma^3_i$ and $\mathcal{B}_p=I-\prod_{i\in p}\sigma^1_i$ are four-body projectors constructed from the products of Pauli operators around each vertex $v$ and plaquette $p$, respectively. Recall that the physical spins are at the edges of the lattice. The state associated with general $\alpha$ and $\beta$ can be obtained from the toric code by the local transformation up to a normalization factor as
\begin{equation}
\ket{\psi}=\prod_v \mathcal{U}_v\ket{\mathrm{TC}},
\end{equation}
where $\mathcal{U}_v=\exp\{h_\mathrm{b}\sigma^3_\mathrm{b}+h_\mathrm{t}\sigma^3_\mathrm{t}+h_\mathrm{bt}\sigma^3_\mathrm{b}\sigma^3_\mathrm{t}\}$ are two-body operators acting vertically on each vertex. Here, the labels $\mathrm{b}$ and $\mathrm{t}$ represent the bottom and top edge of a vertex and $h_\mathrm{b}=\frac{1}{8} \ln \left(\frac{(\alpha
   -1) \alpha }{(\beta -1) \beta }\right)$, $h_\mathrm{t}=\frac{1}{8} \log \left(\frac{\alpha  (\beta -1)}{(\alpha -1)
   \beta }\right)$, $h_\mathrm{bt}=\frac{1}{8} \log
   \left(\frac{\alpha  \beta }{(\alpha -1) (\beta
   -1)}\right)$. The state $\ket{\psi}$ is therefore annihilated by 

\begin{equation}
\widetilde{\mathcal{A}}_v=\prod_{v'}\mathcal{U}_{v'}\mathcal{A}_v\prod_{v'}\mathcal{U}_{v'}^{-1}=\mathcal{A}_v,
\end{equation}
\begin{equation}
\widetilde{\mathcal{B}}_p=\prod_{v'}\mathcal{U}_{v'}\mathcal{B}_p\prod_{v'}\mathcal{U}_{v'}^{-1}, \label{eq:new_annihilated_plumb}
\end{equation}
where in the first equation we use the fact that $\mathcal{A}$ and $\mathcal{U}$ only consist of $\sigma^3$ and thus commute with each other.
The corresponding parent Hamiltonian can be chosen as
\begin{equation}
\widetilde{H}=\sum_v \widetilde{\mathcal{A}}_v^\dagger \widetilde{\mathcal{A}}_v+\sum_p \widetilde{\mathcal{B}}_p^\dagger \widetilde{\mathcal{B}}_p.
\end{equation}
Note that in Eq.~(\ref{eq:new_annihilated_plumb}), only the terms associated with the vertices which sit on the corner of the plaquette p contribute to $\widetilde{\mathcal{B}}_p$, as shown below 
\begin{equation}
\begin{tikzpicture}[baseline=(current  bounding  box.center), scale=1.0]
\crossdot{0}{0.5}
\crossdot{1}{0.5}
\crossdot{0.05}{1.5}
\crossdot{0.95}{1.5}
\crossdot{-0.05}{-0.5}
\crossdot{1.05}{-0.5}
\crossdot{0.5}{0}
\crossdot{0.5}{1}
\crossdot{-0.5}{0}
\crossdot{-0.5}{1}
\crossdot{1.5}{0}
\crossdot{1.5}{1}
\crossdot{-1}{-0.5}
\crossdot{-1}{0.5}
\crossdot{-1}{1.5}
\crossdot{2}{-0.5}
\crossdot{2}{0.5}
\crossdot{2}{1.5}
\draw[very thick, color=red] (0,0) rectangle (1,1);
\draw[very thick, color=blue] (-0.1,-0.8)--(-0.1,1.0);
\draw[very thick, color=blue] (1.1,-0.8)--(1.1,1.0);
\draw[very thick, color=blue] (0.1,1.8)--(0.1,0.0);
\draw[very thick, color=blue] (0.9,1.8)--(0.9,0.0);
\end{tikzpicture}
\;.
\end{equation}
Here the red line denotes one plaquete operator $\mathcal{B}_p$, and the blue lines denote the four $\mathcal{U}_v$ which contribute in Eq.~(\ref{eq:new_annihilated_plumb}). The crosses represent the physical spins, which are at the edge of the lattice. From this, we can see that $\widetilde{\mathcal{B}}_p$ is at most an $8$-body operator. Since $\widetilde{\mathcal{A}}_v=\mathcal{A}_v$ is a $4$-body operator, the parent Hamiltonian of the example Eq.~(\ref{eq:plumbing_example_topo}) is at most $8$-local. This can be compared to a general ``plumbing'' PEPS describing a eight-vertex model, where the parent Hamiltonian is $12$-local~\cite{liu2024topological}.

For the special points where $\beta=1-\alpha$, we found that $h_\mathrm{bt}=0$, thus $\mathcal{U}_v$ is decomposed to a product of two single-site operators. As can be seen in Eq.~(\ref{eq:new_annihilated_plumb}), $\widetilde{\mathcal{B}}_p$ has the same locality as $\mathcal{B}_p$. Therefore, the parent Hamiltonians for those states are at most four-local.

\section{Explicit form of $W$ tensor}\label{app:Wtensor}

Here we report the general expression for a  plumbing tensor to be a DI-PEPS. To this end, the explicit form of $W$ tensor can be obtained by directly substituting Eq.~(\ref{eq:plumbing}) to Eqs.~(\ref{eq:ISOTNS}) and~(\ref{eq:Dual_ISO_TNS}). The parametrization of the resulting $W$ tensor can be written as
\begin{equation}
W_{\mathrm{lb,rt}}=\begin{pmatrix}\cos\alpha\cos\theta_{1}e^{i\phi_{1}} & \cos\alpha\sin\theta_{1}e^{i\phi_{2}} & \sin\alpha\sin\theta_{2}e^{i\phi_{3}} & \sin\alpha\cos\theta_{2}e^{i\phi_{4}}\\
\cos\beta\sin\theta_{5}e^{i\phi_{9}} & \cos\beta\cos\theta_{5}e^{i\phi_{10}} & \sin\beta\cos\theta_{6}e^{i\phi_{11}} & \sin\beta\sin\theta_{6}e^{i\phi_{12}}\\
\sin\alpha\sin\theta_{3}e^{i\phi_{5}} & \sin\alpha\cos\theta_{3}e^{i\phi_{6}} & \cos\alpha\cos\theta_{4}e^{i\phi_{7}} & \cos\alpha\sin\theta_{4}e^{i\phi_{8}}\\
\sin\beta\cos\theta_{7}e^{i\phi_{13}} & \sin\beta\sin\theta_{7}e^{i\phi_{14}} & \cos\beta\sin\theta_{8}e^{i\phi_{15}} & \cos\beta\cos\theta_{8}e^{i\phi_{16}}
\end{pmatrix}.
\end{equation}
with all the angle variables take values in $[0,2\pi)$. Note that the toric code corresponds to a special choice of
the parameters $\theta_{i}=\phi_{j}=0$ for all $i\in\{1,\cdots,8\},j\in\{1,\cdots,16\}$
and $\alpha=\beta=\frac{\pi}{4}$.

\end{document}